\documentclass[12pt,a4paper]{iopart}
\pdfoutput=1
\usepackage{color}
\usepackage{graphicx}
\usepackage{subfigure}
\usepackage{lineno}
\usepackage{bm, amssymb, latexsym,amsmath,verbatim}
\usepackage{iopams}
\usepackage{pdfpages}

\renewcommand{\bs}{\mathbf{s}}
\newcommand{\xx}{\mathbf{x}}
\newcommand{\yy}{\mathbf{y}}
\newcommand{\pphi}{\boldsymbol{\phi}}
\newcommand{\eeta}{\boldsymbol{\eta}}
\newcommand{\xxi}{\boldsymbol{\xi}}
\newcommand{\aalpha}{\boldsymbol{\alpha}}
\renewcommand{\bbeta}{\boldsymbol{\beta}}
\newcommand{\mmu}{\boldsymbol{\mu}}
\newcommand{\perm}[1]{\mathrm{sym}[#1]}
\newcommand{\crossd}[5]{{{#1}^{{\{#2\}_{#3}\{#4\}_{#5}}}}}
\newcommand{\crossdinv}[5]{{{#1}^{{\{#4\}_{#5}\{#2\}_{#3}}}}}

\newcommand{\enh}[1]{\textcolor{blue}{#1}}

\begin{document}

\title{Cooperation, competition and the emergence of criticality in communities of adaptive systems}

\author{Jorge Hidalgo$^{1,2,\ast}$, Jacopo Grilli$^{2,3}$, Samir Suweis$^{2}$, Amos Maritan$^{2}$ and Miguel A. Mu\~noz$^{1,\dag}$} 
\address{$^{1}$Instituto Carlos I de F\'isica Te\'orica y Computacional and Departamento Electromagnetismo y F\'isica de la Materia, Universidad de Granada, 18071 Granada, Spain. 
$^{2}$Dipartimento di Fisica `G. Galilei' and CNISM, INFN, Universit\'a di Padova, Via Marzolo 8, 35131 Padova, Italy. 
$^{3}$Department of Ecology \& Evolution, University of Chicago, 1101 E 57th St, Chicago, IL 60637, USA. 
$^{\ast}$jhidalgo@onsager.ugr.es, 
$^{\dag}$mamunoz@onsager.ugr.es
}

\begin{abstract}
  The hypothesis that living systems can benefit from operating at the vicinity of critical points has gained momentum in recent years. Criticality may confer an optimal balance between too ordered and exceedingly noisy states. Here we present a model, based on information theory and statistical mechanics, illustrating how and why a community of agents aimed at understanding and communicating with each other converges to a globally coherent state in which all individuals are close to an internal critical state, i.e. at the borderline between order and disorder. We study --both analytically and computationally-- the circumstances under which criticality is the best possible outcome of the dynamical process, confirming the convergence to critical points under very generic conditions. Finally, we analyze the effect of cooperation (agents trying to enhance not only their fitness, but also that of other individuals) and competition (agents trying to improve their own fitness and to diminish those of competitors) within our setting. The conclusion is that, while competition fosters criticality, cooperation hinders it and can lead to more ordered or more disordered consensual outcomes.  \end{abstract}

\section{\label{sec:intro}Introduction}

Empirical evidence is mounting that living systems and communities of them might operate at the vicinity of a critical point \cite{Beggs2003,Kauffman08,Peterman2009,chen2012scale,Cavagna2012,Taglia,Kaneko,Chialvo13,Bialek2014,Arcangelis2014,Schuster,Shew2015}.  Criticality, with its concomitant scale invariance, power laws distributions, and extremely large correlations and response \cite{Binney,Lesne} could be a possible source of functional advantages for biological systems\cite{Bialek2011}. In particular, in the brain --one of the flagships of the criticality hypothesis-- it could lead to maximal dynamic ranges, high sensitivity to stimuli, optimal transmission and storage of information, and very diverse dynamical repertoires \cite{Beggs2008,Chialvo10,Kinouchi-Copelli,mora2010maximum,Plenz2013,swarms}.  Different mechanisms and scenarios have been described in the recent literature to explain how such a critical behavior comes about \cite{SOC-Levina,Bonachela1,SOC-Jabonachela,SOC-Millman,Plenz2015}. On the other hand, some authors have argued that apparent criticality could be just an artifact of the attempt to fit overly simplified models to complex and highly heterogeneous systems \cite{Marsili1,Marsili2}.

In a recent work with a fresh perspective ~\cite{our-PNAS}, criticality has been shown to emerge in communities of individuals/agents trying to communicate with each other and creating a collective entity. This approach considers a community of individuals equipped with a (genetic, neural, regulation...) network representing the internal configuration of each individual agent.  The state of each of such networks is controlled by some parameters that completely determine the steady state probability distribution function of internal configurations.  It is assumed that the state of each single agent is defined by such a probability distribution.  Each individual of the community tries to mimic, i.e. to infer or ``understand'', the state of others within a community. With unexpected generality, under this dynamics, the community experiences a drift toward the critical point of the network dynamics, i.e. at the edge between ordered and disordered states.

Remarkably, this emerging criticality entails a large variability among individuals; indeed, in the critical regime, small variations in parameter values are reflected in large state changes. This leads to the somehow surprising conclusion that individuals aiming at understanding each other in the best possible way --thus having the possibility to coordinate their behavior and reactions-- end up exhibiting a large variability. Naively, one could have anticipated that such individuals could have converged to an ``ordered phase'' in which a deterministic and fixed output would be easily predictable or, alternatively, to a ``disordered phase'' in which they all would be essentially random and alike.  However, as shown in \cite{our-PNAS}, such hypothetical solutions turn out to be unstable against fluctuations or noise, and the criticality is the only feasible outcome.

In this paper, we present a variant of the model proposed in~\cite{our-PNAS}, which maintains the same phenomenology, but is formulated in a simpler setting. This allows us to characterize quantitatively the dynamics, being able to describe mathematically both the steady state and the transient under different conditions.  Using this simpler model, we elucidate how and why criticality comes about and investigate what are the roles of cooperation and competition among agents.

The manuscript is structured as follows: in Section \ref{sec:framework} we outline the framework introduced in \cite{our-PNAS} in a simple and concise way.  In Section \ref{sec:model}, we describe a variant of the original model susceptible to mathematical analysis, while in Section \ref{sec:results} we study the attractors of such dynamics by employing a mathematical formalism which allows to map a community of interacting agents into an effective Fokker-Planck equation \cite{Dean}. Finally, in Section \ref{sec:nu} we study a generalization of the model in which individuals can also cooperate or compete.

\section{Mathematical preliminaries}
\label{sec:framework}
Consider a community of $N$ individuals/agents trying to imitate as much as possible the state of each other.  Each of these individuals is characterized by a set of $M$ internal computing units (i.e. nodes of a network) that, for simplicity, we assume to take binary values. Therefore, the internal state of each single agent is fully characterized by the string $\bs = (s_1,...,s_{M})\in\{0,1\}^{M}$. This state changes in time, $\bs=\bs(t)$, obeying some dynamical rules leading to a stationary probability distribution $P(\bs|\xx)$, where the --$d$-dimensional-- variable $\xx=(x_1,...,x_d)$ accounts for all the parameters modulating the dynamics (where $d$ is the dimension of the parameter space), as for instance the strength of couplings between network nodes.  Each individual agent (i.e. its internal state) is completely determined by the value of $\xx$ and, for simplicity, all individuals are assumed to be identical and to obey the same internal dynamical rules except, possibly, for the specific values of their internal parameters. In other words, each agent is characterized by its coordinates in a common parameter space.

Let us emphasize that the framework introduced here consists in a rather sketched idealization of specific biological systems, which aims at generality rather than specificity. However, to provide the reader with some intuition, one could think of a community of bacteria in which individuals are sensitive and responsive to others, and they reconfigure their internal state (as described for instance by their gene regulatory networks, with intricate feedback mechanism between signals and gene state switches) in order to behave similarly to others. This may be relevant in a variety of processes that require of (or benefit from) a collective response at the community level. The main goal of this work is to underline the non-trivial attractors that could emerge out of this type of ``imitation'' game at its relationship with criticality. Certainly, our framework should be carefully adapted to describe specific biological problems.

Before proceeding --and for the sake of self-consistency-- we briefly recall two basic concepts of information theory: the Kullback-Leibler divergence and the Fisher information. Readers familiar with these concepts can safely skip the following two paragraphs.

{\bf The Kullback-Leibler (KL) divergence} \cite{Cover-Thomas} allows for quantifying the difference between two probability distributions. For instance, given the parameter sets $\xx$ and $\yy$ characterizing two probability distribution functions, the KL divergence from the second distribution to the first is defined as \cite{Cover-Thomas}: \begin{equation}
 D(\xx,\yy) = D(P(\cdot|\xx),P(\cdot|\yy))= \left\langle \log{\frac{P(\cdot|\xx)}{P(\cdot|\yy)}} \right\rangle_{\xx} ,
\label{eq:KL}
\end{equation}
where the average $\langle \cdot \rangle_{\xx}$ is taken over $P(\cdot|\xx)$.  In short, Eq. (\ref{eq:KL}) measures the deficit of information when $P(\bs|\yy)$ is used to approximate $P(\bs|\xx)$ \cite{Cover-Thomas}.  Importantly, the KL divergence constitutes a pseudo-distance, as it does not obey the triangle inequality, and is not symmetric as in general $D(\xx,\yy) \neq D(\yy,\xx)$, except in the case in which both distribution are identical (in which case the divergence vanishes).

{\bf The Fisher information (FI)} is a measure of how {\it distinguishable} is a (finite) dataset extracted from a probability distribution from another one obtained with slightly different parameter values.  For example, there could be a region in $\xx$ space in which $P(\bs|\xx)$ are mostly invariant as we change $\xx$, while in another regions the distribution could be highly sensitive to parameter changes. The FI is defined as \cite{Cover-Thomas,Marsili1}: \begin{equation}
  \chi_{\alpha\beta}(\xx)= \left\langle \frac{\partial \log P(\cdot|\xx)}{\partial x_\alpha} \frac{\partial \log P(\cdot|\xx)}{\partial x_\beta} \right\rangle_{\xx},
  \label{eq:FI}
\end{equation}
for $\alpha,\beta=1,...,d$. The determinant $\det(\chi(\xx))$ is a measure of the density of distinguishable distributions in parametric space (as a function of $\xx$) \cite{Balasubramanian1997,Marsili1}. In particular, if $\det(\chi(\xx))$ peaks at $\xx^*$, this corresponds to a
region in the space of parameters $\xx$ in which distributions are highly sensitive to changes of the parameters.  Thus, it is not surprising that the FI exhibits a peak at critical points \cite{Marsili1}. 

To illustrate these concepts, without loss of generality, we can parametrize the internal probability distribution of agents as: \begin{equation}
 P(\bs|\xx) = \frac{\exp\left(-\xx \cdot \pphi(\bs) \right)}{\sum_{\bs'}\exp\left(- \xx \cdot \pphi(\bs') \right)}
\label{eq:parametrization}
\end{equation}
where $\pphi=(\phi_1,...,\phi_d)$ are the so-called ``observable'' functions (i.e. functions of the internal configuration) and $\xx\cdot\pphi = \sum_{\alpha=1}^d x_\alpha \phi_\alpha$. If, in particular, $\phi_1(\bs)=\sum_{i<j}^{M}s_i s_j/M$, the internal state of each agent corresponds to a mean-field Ising model at some temperature \cite{Binney}.  For such a parametrization, Eq. (\ref{eq:parametrization}), the FI is $\chi_{\alpha\beta}(\xx)=-\partial \langle \phi_{\alpha} \rangle_{\xx}/\partial x_\beta= \langle \phi_\alpha\phi_\beta \rangle_\xx -\langle \phi_\alpha \rangle_\xx \langle \phi_\beta \rangle_\xx$, which corresponds to the generalized susceptibility, that is well-known to diverge at critical points. Consequently, in terms of information theory, most distinguishable patterns are concentrated around the critical point \cite{Marsili1}.

In what follows, we mainly work with two different parametrizations. The simplest one corresponds to the (zero-field) Ising mean-field model \cite{Binney}: \begin{equation}
  P(\bs|x_1)\propto\exp\left(x_1\sum_{i<j}^{M}s_i s_j/{M} \right),
 \label{eq:ising_1d}
\end{equation}
which has the advantage of having just only one free parameter $\xx=x_1$ (usually interpreted as the inverse temperature) and 
has a FI peaked at $x_1^*\simeq1$, which diverges at $x_1^c=1$ in the (thermodynamic) limit, $M\rightarrow\infty$.
Similarly, we also consider the mean-field Ising model with an external field (which has two parameters):
\begin{equation}
P(\bs|x_1,x_2)\propto\exp\left(x_1\sum_{i<j}^{M}s_i s_j/{M} + x_2\sum_{i}^{M}s_i\right),
\label{eq:ising_2d}
\end{equation}
which has a FI that diverges as $M\rightarrow\infty$ at $(x_1\geq1,x_2=0)$. This corresponds to i) a line of first order phase transitions for $x_2=0$ and $x_1\geq1$ and ii) a second order phase transition at $x_2=0$ and $x_1=1$. As it will become clear in what follows, we are more interested in the second case as, for any finite system size $M$, the maximum of the determinant of the FI is closer to $\xx^c=(1,0)$ (i.e. the critical point of the second-order phase transition) than to any other point along the discontinuous phase transition line. We use the notation $\xx^c$ for the critical point -- defined in the thermodynamic limit $M\rightarrow\infty$ -- while $\xx^*(M)$ is used for the maximum of $\det(\chi(\xx))$ in parameter space for any given size $M$.

\section{Model}
\label{sec:model}
\begin{figure}
    \centering\includegraphics[width=0.7\columnwidth]{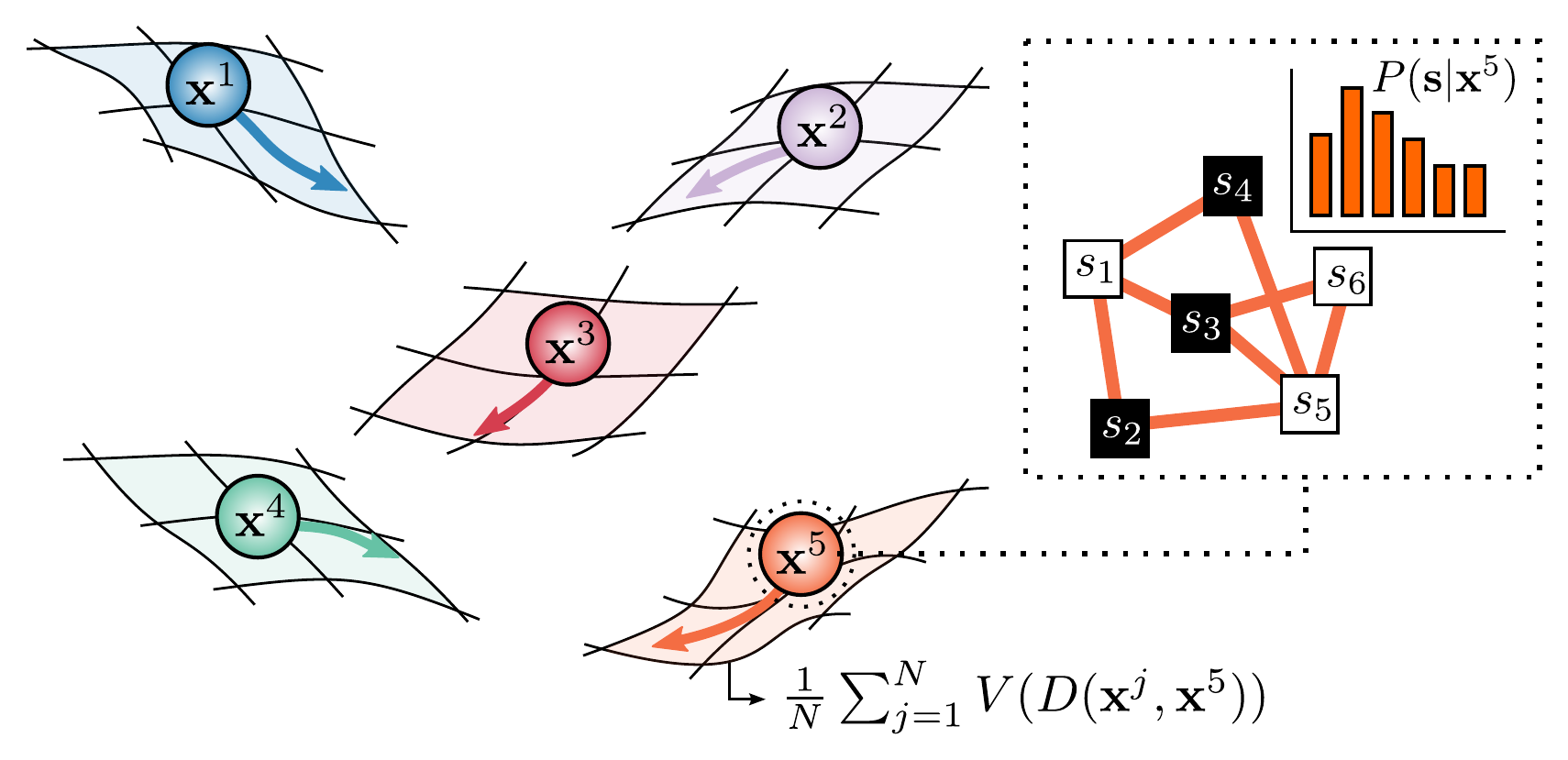} \caption{A community of individuals adapts to increase the information each one is able to infer from the others. 
    Each individual $i$ is composed of $M$ internal binary units; its internal state is represented by the probability distribution $P(\bs|\xx^i)$ for each configuration $\bs=(s_1,...,s_{M})$. The free parameter $\xx^i$ controls the internal dynamics and all individuals obey the same dynamical rules (some type of probability distribution function, but possibly with different parameter values).  At each time, individual $i$ changes its parameter $\xx^i$ in the direction that minimizes its average information-deficit respect to the community, $\sum_{j=1}^N V(D(\xx^j,\xx^i))/N$, where $D(\xx^j,\xx^i)$ is the Kullback-Leibler divergence (Eq. (\ref{eq:KL})) from $P(\bs|\xx^i)$ to $P(\bs|\xx^j)$, and $V(D)$ is some increasing (pseudo-potential) function of $D$. Additionally, there is some randomness in the dynamical process. The focus is in the adaptation of the whole community in the space of parameters $\xx$; does it converge, and if so, where?} \label{fig:model} \end{figure}

We introduce a model, which is identical in spirit to the co-evolutionary model in \cite{our-PNAS} but which, in contrast, enables for analytical understanding. Here we discuss both the original and the modified models, stressing their similitudes and differences. Both models consist of a community of $N$ individuals aiming at having an internal state/distribution as similar as possible to the others --as quantified by their KL divergences to other individuals-- i.e. aiming at optimizing the information they have from the rest of the community, i.e. at minimizing its information deficit (see Fig. \ref{fig:model}). In both cases, individual agents are characterized by an internal state (probability distribution function) parametrized as described in the previous section. All individuals are identical in principle, but they may differ in their parameter values.

\subsection*{Original co-evolutionary model}
In the original model (see ``coevolutionary model'' in \cite{our-PNAS}) the dynamics proceeded by randomly selecting at each time step a pair of agents; given that the KL divergence is not symmetric, one of the two agents has a larger ``fitness'' (i.e. it infers better the state of the other than the other way around) and thus, as a consequence of this informational advantage, it has a larger probability of generating progeny.  Reversely, the less fit agent is more likely to die and be removed from the community, while the fittest one generates an offspring that inherits its parameter values -- with some small variability-- from it.  This dynamical process is iterated in time, defining in this way a genetic algorithm. After sufficiently long time the stable output of this dynamics turned out to be that the community of agents evolved to have intrinsic parameters located around the corresponding critical point.

Results in \cite{our-PNAS} were mostly computational, even if some heuristic understanding was also provided.  Analytical progress was hindered by the intrinsic difficulty of dealing mathematically with the genetic algorithm as defined above.  For this reason, our first goal in this paper is to construct a version of the original model that, while leaving intact the main phenomenology allows for analytical treatment.

\subsection*{Novel adaptive model}
With this motivation, we propose to analyze a slightly different model. This is not an evolutionary one, in the sense that agents do not die nor reproduce. Instead it is an adaptive one in which agents slightly change their parameters trying to enhance their fitness. Thus, we model each agent $i$ by its position in parameter space ${\xx}^i $; it experiences an \emph{adaptive force} which is a function of its information-deficit respect to the other agents, plus some stochastic noise and that can be written as the derivative of some (pseudo)potential, $V$. In particular, the adaptive model is defined by means of the set of Langevin equations \begin{equation}
  \dot{\xx}^i =  \frac{1}{N} \sum_{j=1}^N \mathbf{F}(\xx^j,\xx^i)+\sqrt{2T} \eeta^i(t),
  \label{eq:langevin}
\end{equation}
for $i=1,...,N$, 
where 
\begin{equation}
 \mathbf{F}(\xx^j,\xx^i) = -\nabla_{\xx^i}  V(D(\xx^j,\xx^i))
 \label{eq:drift}
\end{equation}
is the force that an individual agent $j$ produces on an agent $i$, which depends solely on the KL divergence from the second to the first. The first term in the r.h.s. of Eq. (\ref{eq:langevin}) is the averaged force acting upon agent $i$, which points in the direction of the potential gradient. Observe that, in the most general case, $V$ is a pseudo-potential and not a true potential, because, owing to the asymmetry of the KL divergence, $D$, the forces in Eq. (\ref{eq:langevin}) do not obey the potentiality (Schwarz) condition.  Still, even in such cases, the force can be defined as the derivative of such a pseudo-potential.  The second term in Eq. (\ref{eq:langevin}) represents a thermal noise --which is an ineluctably present ingredient of any biological process-- and that as we will see plays an important role in the dynamical process (getting rid of unstable solutions). In particular, $\sqrt{2T}$ modulates the noise amplitude, and $\eeta^i$ is a white noise with $\langle \eta^i_\alpha(t) \rangle=0$ and $\langle \eta^i_\alpha(t) \eta^j_\beta (t') \rangle = \delta^{ij} \delta_{\alpha\beta} \delta(t-t')$ ($i,j=1,...,N$, and $\alpha,\beta=1,...,d$).

To complete the model definition we still need to specify the functional form of the pseudo-potential $V(D(\xx^j,\xx^i))$. The main constraint we need to impose is that $V$ has to increase monotonically with $D$ (i.e. $V'(D)\geq0 ~~\forall D\geq0$) in order to guarantee that the dynamical process, which converges to minima of $V$, leads also to minimal values of the KL divergence. In this way, at each time step, each agent $i$ changes its parameter, $\xx^i$, in the direction of the gradient of the averaged $V$, with some added stochasticity. Here --for reasons explained in detail in \ref{sec:concavity}-- we restrict our analysis to the case of a convex potential $V''(D)>0$.  Having specified the shape of $V$ and thus the full model, the question is: where does the community evolve to as a consequence of this information-based dynamics? In the next section we derive a mathematical approach to compute the attractors of the dynamics in the space of parameters $\xx$ for different choices of the (pseudo)potential $V(D)$.

\section{Analytical and computational results}
\label{sec:results}

\subsection{Community dynamics} \label{sec:dean} We are interested in the evolution of the distribution of agents in parameter space, defined as: \begin{equation}
  \rho(\xx,t) = \frac{1}{N} \sum_{i=1}^N \delta(\xx^i(t)-\xx).
\end{equation}
Luckily enough, there exists an analytical method to map a set of particle-like agents (interacting among themselves by means of 
Langevin dynamics) into an equation for the probability distribution of their positions in the space in which they move.
Such a method was proposed by Dean in \cite{Dean} (an earlier reference to this equation can be found in the work of Kawasaki \cite{Kawasaki}), and it allows us to derive an equation for $\partial_t \rho(\xx,t)$ from Eq. (\ref{eq:langevin}). This calculation can be simply understood as the change of variables $\{\xx^i\}\rightarrow\rho$ using the Ito's calculus  (see \ref{sec:Dean} or ref. \cite{Dean} for the explicit derivation), leading to the following equation for the probability distribution:
\begin{equation}
  \partial_t \rho(\xx,t) =  
 -\nabla_\xx \cdot \left( \rho(\xx,t) \displaystyle \int d\yy \rho(\yy,t) \mathbf{F}(\yy,\xx) \right) + T \nabla^2_{\xx}\rho(\xx,t) + \sqrt{\frac{2T}{N}} \nabla_\xx \cdot \left(\sqrt{\rho(\xx,t)} \xxi(\xx,t) \right)
  \label{eq:dean}
\end{equation}
where $\xxi(\xx,t)$ is a new Gaussian noise with zero mean and correlation $\langle \xi_\alpha(\xx,t)\xi_\beta(\yy,t') \rangle = \delta_{\alpha\beta} \delta(\xx-\yy)\delta(t-t')$ ($\alpha,\beta=1,...,d$), interpreted in the Ito's sense \cite{Gardiner}.
Observe that, in the limit $N\rightarrow\infty$, Eq. (\ref{eq:dean}) becomes deterministic, and we find an equation for $\rho(\xx,t)$, which, roughly speaking, is a sort of non-linear Fokker-Planck equation \cite{Gardiner} in which the drift term depends on the distribution $\rho$ itself.

Observe that Eq. (\ref{eq:dean}) synthesizes the evolution of the community in a single dynamical equation. Before proceeding, some remarks are in order: (i) the global gradient ensures the conservation of the total probability (no individuals are created or annihilated); (ii) the diffusive term comes from fluctuations in the dynamics (controlled by $T$); (iii) at any region where there are no individuals (i.e., where $\rho=0$), the distribution does not change (both the deterministic and the stochastic terms vanish; i.e. it is an absorbing state/region \cite{Marro-Dickman}).

We can infer the shape of $\rho$ from Eq. (\ref{eq:dean}); in the case in which the drift force is small (which occurs, for instance, when individuals are very close to each other, as expected if the dynamics is cohesive), the r.h.s. of Eq. (\ref{eq:dean}) becomes negligible, and $\rho$ is expected to look like a Gaussian distribution. Also, for a narrow distribution, we can anticipate that, if we expand the drift term (Eq. (\ref{eq:drift})) in Taylor series around the centroid of the community, first order contributions will cancel after integrating in the community (as individuals approach each other, the overall force almost vanishes), and second order terms --controlling concavity/convexity of $V$-- may play an important role of the dynamics (see \ref{sec:concavity} for further details). Anyway, a detailed calculation needs to be performed in order to clarify these questions for each specific case.

Employing Eq. (\ref{eq:dean}) it is possible to derive a set of deterministic equations for the evolution of 
its moments for $N\rightarrow\infty$. The mean $\mmu(t)= \int d\xx \rho(\xx,t) \xx$ can be obtained multiplying Eq. (\ref{eq:dean}) by $\xx$ and integrating by parts:
\begin{equation}
 \dot \mu_{\alpha} = \displaystyle \int d\xx \rho(\xx,t) \displaystyle \int d\yy \rho(\yy,t) F_\alpha (\yy,\xx),
 \label{eq:mu1}
\end{equation}
where $\alpha=1,...,d$. Similarly for the covariance matrix, defined as $K_{\alpha\beta}(t) =\int d\xx \rho(\xx,t) \left(x_\alpha - \mu_\alpha(t)\right)\left(x_\beta-\mu_\beta(t)\right)$ ($\alpha,\beta=1,...,d$), we find:
\begin{equation}
 \dot K_{\alpha\beta} = 2T\delta_{\alpha\beta}+\sum_{\gamma,\epsilon=1}^d (\delta_{\alpha\gamma}\delta_{\beta\epsilon}+\delta_{\alpha\epsilon}\delta_{\beta\gamma})\displaystyle \int d\xx \rho(\xx,t) (x_\gamma-\mu_\gamma) \displaystyle \int d\yy \rho(\yy,t) F_\epsilon (\yy,\xx).
  \label{eq:K1}
\end{equation}
These equations cannot be integrated as the evolution of the mean and covariance matrix still depends on the whole distribution $\rho$, i.e. they are not a closed set of equations. However, in the next Section we derive an approximation allowing us to circumvent this difficulty and characterize the evolution of the community by the mean and covariance matrix of the distribution in parameter space.

\subsection{Explicit equations for the evolution of the community}
\label{sec:expansion}
Here we develop an approximate scheme giving rise to a closed set of  explicit equations for the evolution of the mean and covariance matrix of parameters in the community.  To this end, we first expand  $\mathbf{F}(\yy,\xx)$ around $(\yy,\xx)=(\mmu(t),\mmu(t))$ in Eqs. (\ref{eq:mu1}) and (\ref{eq:K1}), and integrate over $\xx$ and $\yy$ (see Supplementary material for a detailed derivation); the result is then implicitly given as a function of moments of $\rho$. This expansion can be truncated at some level of approximation; the first contributing terms are (see Supplementary material):
\begin{eqnarray}
\dot \mu_\alpha\  &=&  \frac{V''(0)}{8} \sum_{\beta,\gamma,\epsilon,\theta=1}^d \left( \chi_{\alpha\beta}(\mmu) \chi_{\gamma\epsilon\theta}(\mmu) + \chi_{\alpha\gamma}(\mmu) \chi_{\beta\epsilon\theta}(\mmu)+\right.
\nonumber\\
&& \left. \chi_{\alpha\epsilon}(\mmu) \chi_{\beta\gamma\theta}(\mmu)+\chi_{\alpha\theta}(\mmu) \chi_{\beta\gamma\epsilon}(\mmu) \right) \left(K_{\beta\gamma}K_{\epsilon\theta} + \frac{K_{\beta\gamma\epsilon\theta}}{3} \right)
\label{eq:firstmom1}\\
 \dot K_{\alpha\beta} &=& 2T\delta_{\alpha\beta} - V'(0) 
\sum_{\gamma=1}^d \left(\chi_{\alpha\gamma}(\mmu) K_{\beta\gamma}+\chi_{\beta\gamma}(\mmu) K_{\alpha\gamma}\right)
\label{eq:secondmom1}
\end{eqnarray}
where $\mathbf{\chi_{\alpha\beta}}$ is the susceptibility as defined in Eq. (\ref{eq:FI}), $\chi_{\alpha\beta\gamma}(\mmu)=\left.\partial_{x_\alpha}\chi_{\beta\gamma}(\xx)\right|_{\xx=\mmu}$, and $K_{\alpha\beta\gamma\delta}$ is the $4-th$ moment of the distribution of parameters, $\rho(\xx,t)$.  For the case in which $V$ has a local minimum at $D=0$ (i.e., $V'(0)=0$, $V''(0)>0$), the second term in Eq. (\ref{eq:secondmom1}) vanishes, and one needs to keep further terms in the expansion, obtaining a slightly more complex expression (see \ref{sec:v0}).

\begin{figure}
 \centering\includegraphics[width=0.7\columnwidth]{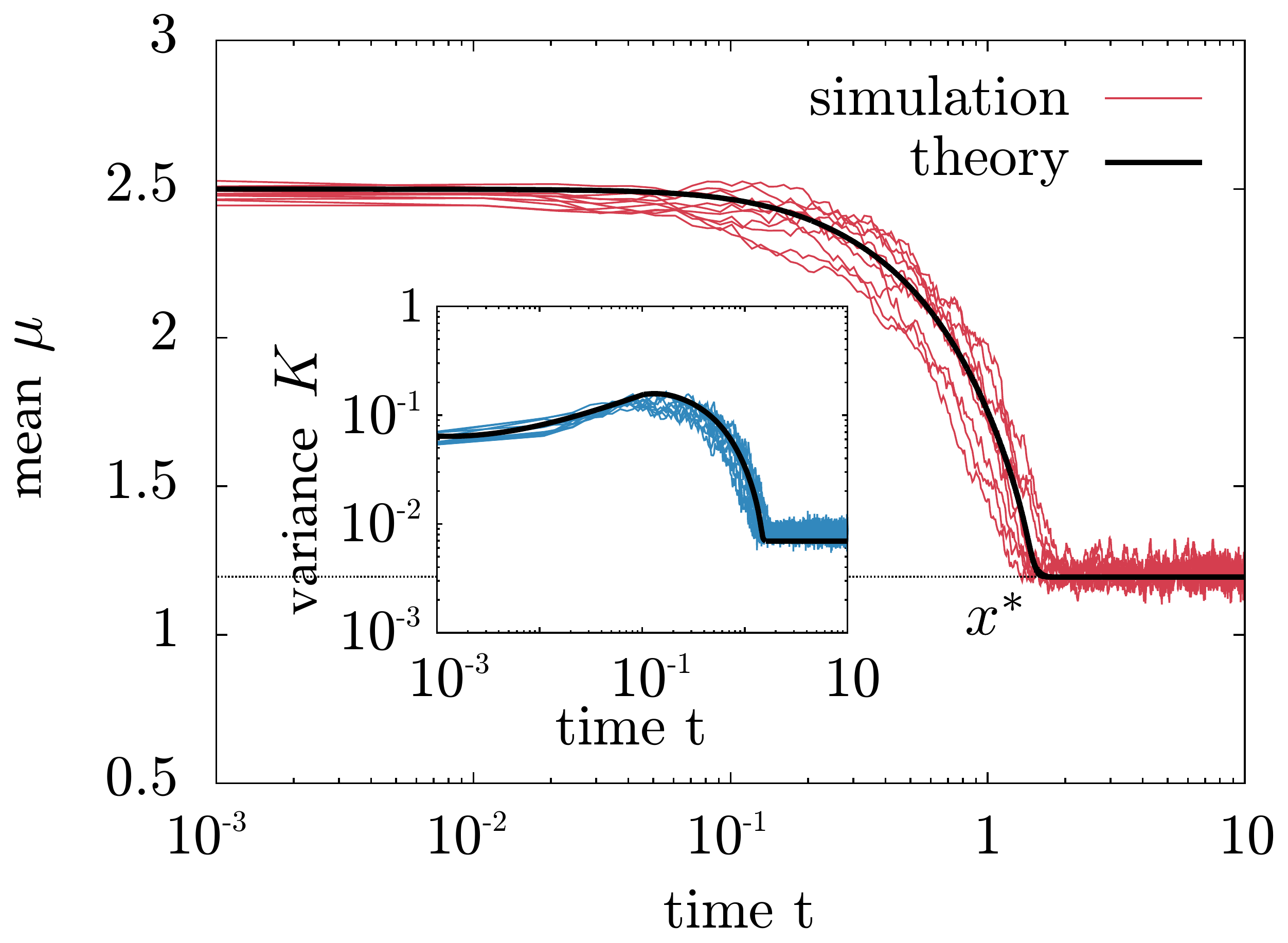}
 \caption{Mean and variance of the distribution of parameter values in a community of individuals adapting according to Eq. (\ref{eq:langevin}), with the one-dimensional Ising-mean internal parametrization Eq. (\ref{eq:ising_1d}). In this example, the local potential is $V(D)=D^2/2$. Colored fluctuating lines represent $10$ independent computer simulations of the dynamics, the solid black curves correspond to the numerical integration of the theoretical prediction for $N\rightarrow\infty$ given by Eqs. \ref{eq:firstmom1-1dim} and \ref{eq:secondmom2-1dim}. Parameters: number of individuals $N=100$, number of internal nodes $M=100$, noise amplitude $T=1$. The maximum of the Fisher Information is located at $x^*\simeq 1.2$ for $M=100$, (approaching to $x^c=1$ in the thermodynamic limit).}
 \label{fig:1dim}
\end{figure}

\begin{figure}
 \centering\includegraphics[width=0.7\columnwidth]{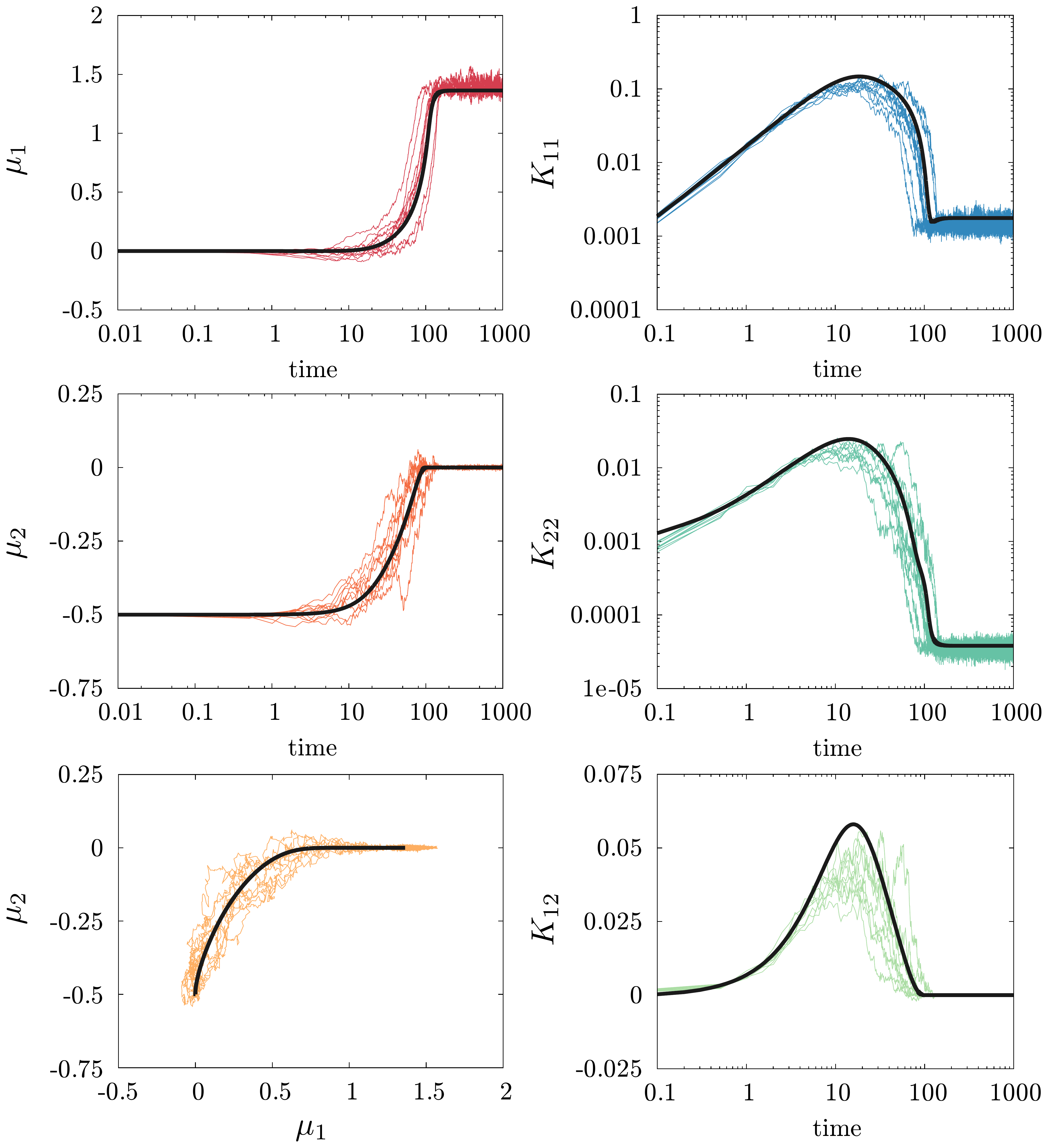}
 \caption{Mean and covariance matrix of parameters in a community of individuals evolving according to Eq. (\ref{eq:langevin}), with the two-dimensional Ising mean-field internal parametrization, Eq. (\ref{eq:ising_2d}). As in Fig. \ref{fig:1dim}, the local potential is $V(D)=D^2/2$. Colored fluctuating lines correspond to 10 independent realizations of the dynamics, and the solid black ones to the numerical integration of the theoretical prediction for $N\rightarrow\infty$, given by eqs. \ref{eq:firstmom1} and \ref{eq:secondmom2} (using the Gaussian approximation $K_{\alpha\beta\gamma\delta}\simeq K_{\alpha\beta}K_{\gamma\delta} + K_{\alpha\gamma} K_{\beta\delta} + K_{\alpha\delta} K_{\beta\gamma}$). Parameters: number of individuals $N=100$, number of internal nodes $M=100$, noise amplitude $T=10^{-2}$. The maximum of determinant of the Fisher Information is located at $(x_1^*,x_2)\simeq (1.3,0)$ for $M=100$, (approaching to $(x_1^c,x_2^c)=(1,0)$ in the thermodynamic limit).}
 \label{fig:2dim}
\end{figure}

Equations (\ref{eq:firstmom1}) and (\ref{eq:secondmom1}) cannot be integrated, as we still need additional equations for the 4-th moment $K_{\alpha\beta\gamma\delta}$. We circumvent this problem by approximating the 4-th moment in terms of the second ones (moment closure). Thus we assume that $\rho$ is approximately Gaussian.  To justify this ansatz we can argue that, for small noise variance $T$, individuals stay close to each other. As revealed by eq. \ref{eq:firstmom1}, the effect of the drift is weaker for narrower distributions, and the dynamics is essentially diffusive (at least locally), which leads to a Gaussian distribution. Furthermore, we have corroborated that the results obtained from this approach match well numerical simulations for small values of $T$, while the approximation breaks down for larger values. Therefore, we simply take $K_{\alpha\beta\gamma\delta} \simeq K_{\alpha\beta}K_{\gamma\delta} + K_{\alpha\gamma} K_{\beta\delta} + K_{\alpha\delta} K_{\beta\gamma}$, obtaining closed equations for the first and second moments.  In particular, for the case of just one parameter ($d=1$) we find: \begin{eqnarray}
 \dot \mu  &=& V''(0) \chi(\mu) \partial_\mu \chi(\mu) K^2 \label{eq:firstmom1-1dim}\\
 \dot K &=& 2T-2V'(0) \chi(\mu) K\label{eq:secondmom1-1dim}
\end{eqnarray}
or a slightly different second equation if $V'(0)=0$ (\ref{sec:v0}).
From this set of equations, we can see that: 
\begin{itemize}
\item The centroid of the community, $\mu$, follows the direction of the gradient of the FI, and the sign of $V''(0)$ controls the direction in which individuals move: while the community follows the direction which maximizes the FI for $V''(0)>0$, the opposite occurs for $V''(0)<0$, i.e., individuals follow the direction which minimizes it (see \ref{sec:concavity}). If the FI experiences an abrupt change of its shape, this should be translated into an abrupt change of the dynamics (see, for instance, how the community rapidly goes from the transient to the stationary regime in Figs. \ref{fig:1dim} and \ref{fig:2dim}).

\item The variance of the distribution, $K$, changes owing to two opposite mechanisms: it increases owing to fluctuations (controlled by $T$), while it shrinks as the result of interactions (faster in regions with higher FI). Therefore, if the FI diverges in the thermodynamic limit, all individuals should be concentrated \emph{at} the critical point.

\item In a noiseless scenario ($T=0$), the variance of the distribution vanishes in the stationary limit, $K(t\rightarrow\infty)=0$ (i.e. all individuals have the same parameter values), and the dynamics can stop before the community has reached the global maximum of the FI. Noise destabilizes these kind of attractors, and the only stable solution is the critical point.
 \end{itemize}
 Imposing stationarity in the previous equations, it is straightforward to derive the steady-state distribution, which is centered at the maximum of the FI, $x^*$, and has variance $K^*=\frac{T}{V'(0)\chi(x^*)}$ (see \ref{sec:v0} for the case $V'(0)=0$). Thus, as $\chi$ diverges for $M\longrightarrow\infty$, all individuals will locate exactly \emph{at} the critical point, $x^c$, in the thermodynamic limit, and very close to it for large but finite internal system sizes.

 Fig. \ref{fig:1dim} represents the mean and variance of parameters of 10 independent computer simulations of the dynamics (Eq. (\ref{eq:langevin})), in a community of $N=100$ individuals with internal parametrization of Eq. (\ref{eq:ising_1d}), for $V(D)=D^2/2$ (the potential has a minimum at $D=0$). Solid dark lines represent the numerical integration of our theoretical prediction for $V'(0)=0$ (see Eqs. (\ref{eq:firstmom1-1dim}) and (\ref{eq:secondmom2-1dim}) in the \ref{sec:v0}). As expected, individuals are attracted towards the maximum of the FI, represented by the solid line at $x^*\simeq1.19$. Fig. \ref{fig:1dim} illustrates the perfect agreement of the theoretical approach with the numerical integration of the dynamics. Let us remark that deviations stem from the finite number of individuals $N$ in computer simulations. Analogous simulations with $V'(0)>0$ are shown in \ref{sec:concavity}, revealing again an excellent agreement between simulation an theory.

Let us remark that the analytical calculation becomes much more involved for the case of more parameters, $d\geq2$, but our simulations (see below) confirm that the same phenomenology --i.e. convergence towards criticality-- emerges. Fig. \ref{fig:2dim} represents the mean and covariance matrix in time (obtained via numerical integration of Eq. (\ref{eq:langevin})) for the two-parameter Ising mean-field-like parametrization, Eq. (\ref{eq:ising_2d}), in a community of $N=100$ individuals and $V(D)=D^2/2$. Solid dark lines represent the numerical integration of eqs. \ref{eq:firstmom1} and \ref{eq:secondmom2} (using the Gaussian approximation). 
An interesting observation is that, as in the one-dimensional case, the community evolves near the critical point of the second-order phase transition (i.e. to the global maximum of the determinant of the FI) rather than along the first-order phase transition line.
In this case, the exact attractor (for finite internal system sizes, $M$) cannot be computed, as the stationary solution of Eq. (\ref{eq:firstmom1}) becomes harder to solve for $d>1$. Certainly, for a finite internal system size, $M=100$, the attractor $\mmu(t\rightarrow\infty)=(1.35,0)$ slightly differs from $\xx^*=\arg\max_{\xx}\det(\chi(\xx))=(1.27,0)$ for $M=100$, but both approach to the critical point, $\xx^c=(1,0)$, in the thermodynamic limit.

\section{Generalization of the model: cooperation and competition}
\label{sec:nu}
Let us now generalize the dynamics to embrace the following mechanisms, commonly studied in evolutionary game theory \cite{Nowak}:
\begin{itemize}
\item {\bf Neutralism}: each individual tends to minimize its information-deficit regardless of others (already analyzed case).

\item {\bf Cooperation}: each individual tends to minimize its information-deficit but also, simultaneously, that of all the other individuals in the community.

\item {\bf Competition}: each individual tries to minimize its information-deficit, and, at the same time, tries to 
  maximize the information-deficit of the others.
\end{itemize} 
To account for all these possibilities, we redefine the potential between individuals $i$ and $j$ as a linear combination of their respective information-deficits: \begin{equation}
 V^\nu(\xx^j,\xx^i) = V(D(\xx^j,\xx^i)) + (2\nu-1)V(D(\xx^i,\xx^j)),
 \label{eq:potential-nu}
\end{equation}
where the parameter $\nu\in[0,1]$ tunes the interaction from competition ($\nu=0$) to cooperation ($\nu=1$), including the intermediate neutral case ($\nu=1/2$).  The generalized dynamics is simply obtained by substituting $V(D(\xx^j,\xx^i))$ with $ V^\nu(\xx^j,\xx^i)$ in Eq. (\ref{eq:langevin}) and Eq. (\ref{eq:drift}).  We call $\nu$ the ``symmetry coefficient'', as $V^{\nu=0}$ and $V^{\nu=1}$ are anti-symmetric and symmetric functions, respectively, under the exchange $i \leftrightarrow j$.

We can perform a similar expansion to that carried in Section \ref{sec:expansion} for the mean and covariance matrix of the distribution of parameters (see Supporting Material). The first non-vanishing terms are:
\begin{eqnarray}
\dot \mu_\alpha(t) &=& (1-2\nu) V'(0) \sum_{\beta,\gamma=1}^d \chi_{\alpha\beta\gamma}(\mmu) K_{\beta\gamma}
\label{eq:firstmom1-nu} \\
\dot K_{\alpha\beta}(t) &=& 2T\delta_{\alpha\beta} - 2\nu V'(0) 
\sum_{\gamma=1}^d \left(\chi_{\alpha\gamma}(\mmu) K_{\beta\gamma}+\chi_{\beta\gamma}(\mmu) K_{\alpha\gamma}\right).\label{eq:secondmom1-nu}
\end{eqnarray}
Notice that Eq. (\ref{eq:firstmom1-nu}) becomes much simpler than its counterpart in the neutral case, $\nu=1/2$, Eq. (\ref{eq:firstmom1}). Further terms in the expansion have to be computed in Eq. (\ref{eq:firstmom1-nu}) for the case $\nu=1/2$ or $V'(0)=0$, and similarly for Eq. (\ref{eq:secondmom1-nu}) for $\nu=0$ (see Supporting material).

Some remarks are in order: (i) Eq. (\ref{eq:firstmom1-nu}) tells us that the community moves in the direction of $\nabla(\sum_{\beta\gamma}\chi_{\beta\gamma}(\mmu(t)) K_{\beta\gamma}(t))$, where the sign is given by $(1-2\nu)$. Therefore, agents move toward the critical point for any $\nu<1/2$, while they move in the opposite direction for any $\nu>1/2$.  (ii) Eq. (\ref{eq:firstmom1-nu}) does not depend on the sign of $V''(0)$, but on the sign of $V'(0)$, which is always positive as $V$ is an increasing function of $D$ (observe also that, the previously discussed issue about the concavity/convexity has no relevance for $\nu\neq 1/2$).  Summarizing, individuals move to the critical point for any ``mostly'' antisymmetric potential ($\nu<1/2$), while they escape from it for any ``mostly symmetric'' potential ($\nu>1/2$). The case $\nu=1/2$ represents a borderline case, which has to be analyzed --as shown before-- keeping higher-order terms in the expansion.

\begin{figure}
\centering\includegraphics[width=0.7\columnwidth]{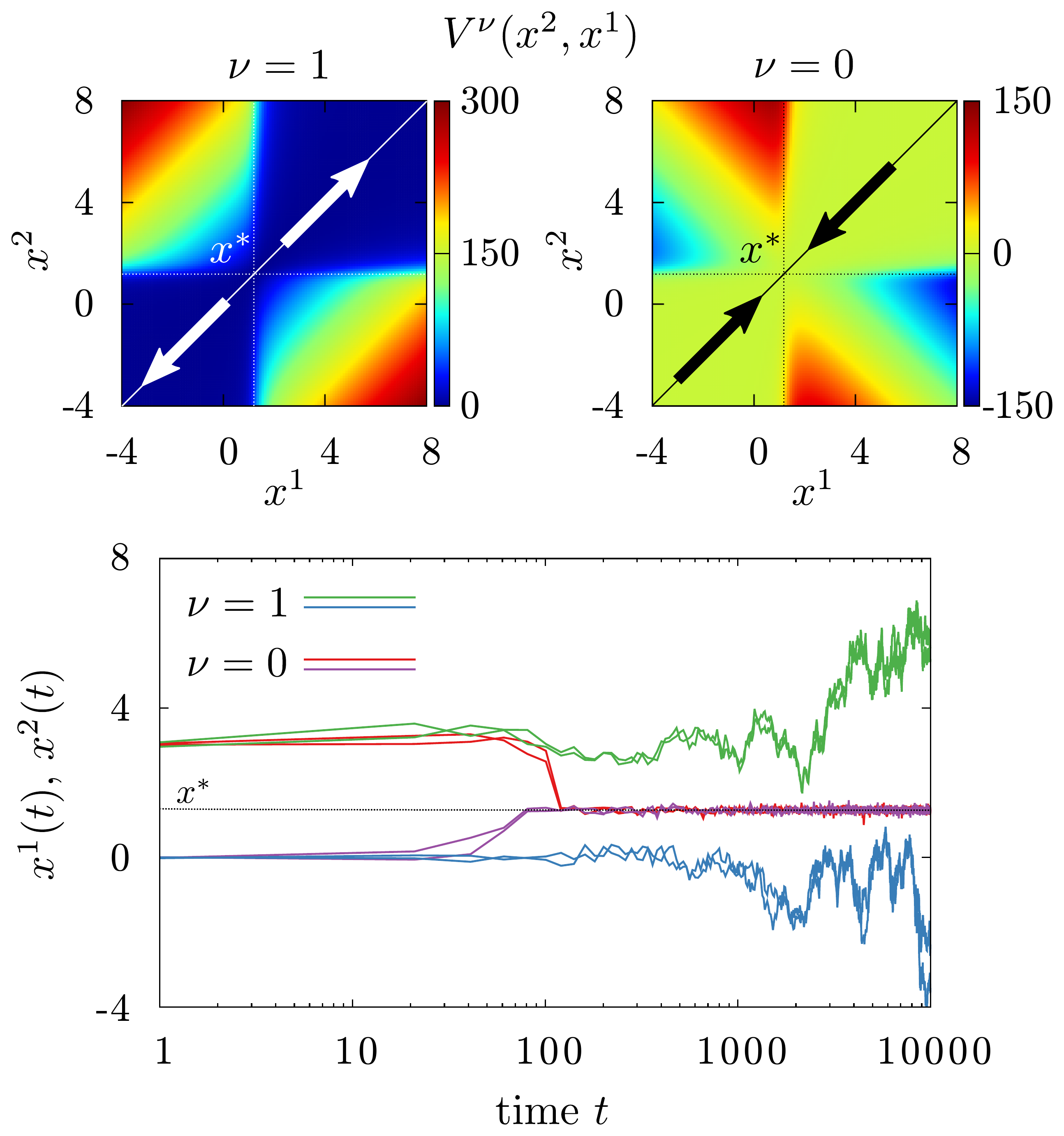} \caption{{\it Top:} Local potential $V^{\nu}(x_2,x_1)$ in the case of cooperation ($\nu=1$, Left) and of competition ($\nu=0$, Right). The first is symmetric under the exchange of $1\leftrightarrow2$, while the second is antisymmetric. Once individuals approach each other, arrows indicate the region in which the centroid is expected to move.  {\it Bottom:} Two individuals evolve their parameters $x^1(t)$ and $x^2(t)$ with the dynamics given by of Eq. (\ref{eq:langevin}) (substituting $V(D(\xx^j,\xx^i))$ with $V^\nu(\xx^j,\xx^i)$). For both $\nu=1$ and $\nu=0$, we have plotted the trajectories of each parameter (using the same color-code) starting from two different initial conditions ($x^1=x^2=0$ and $x^1=x^2=3$, respectively). Simulations illustrate that individuals scape towards regions of lower FI (away from the critical point) in the cooperative case, whereas the opposite occurs in the case of competition, in which they end at the region of highest FI, i.e., the critical point.  In all cases, we have implemented the dynamics using the Ising mean-field like parametrization of Eq. (\ref{eq:ising_1d}) and $V(D)=D$, $M=50$ and $T=10^{-4}$.}
\label{fig:potential}
\end{figure}

Fig. \ref{fig:potential} (Top-Left) illustrates the generalized potential for the case of just two individuals, in the symmetrical case, $V^{\nu=1}(x^1,x^2)$, with the parametrization of Eq. (\ref{eq:ising_1d}) in the linear case $V(D)=D$. Observe that in this case the local potential $V^{\nu=1}(x_1,x_2)=D(x_1,x_2)+D(x_2,x_1)$ is identical to the Jensen-Shannon entropy \cite{Cover-Thomas}.  As the figure shows, there is a flat region in the space of parameters along the principal diagonal.  As near the center, larger potential values (lighter colors) are observed, one can expect that the potential is minimal at $x_1=x_2 \rightarrow\pm\infty$, suggesting that agents should move away from criticality toward highly ordered or highly disordered states. Indeed, this is confirmed by Fig. \ref{fig:potential} (Bottom) which shows simulation results for this case and reveals that agents tend to scape toward regions of low FI, i.e, away from the critical point (similar results are obtained for larger values of $N$; not shown).

The top-right panel of Fig. \ref{fig:potential} illustrates the opposite scenario, i.e. the anti-symmetrical case, $\nu=0$, for the same parametrization (Eq. (\ref{eq:ising_1d})) and $V(D)=D$. As before, we can see a flat region in the space of parameters for $x_1=x_2\rightarrow\pm\infty$ (i.e., away from the critical point).  However, there exists a region of lower information-deficit for one individual (blue region), which is simultaneously maximal for the other one (red region).  Therefore, both individuals want to be close enough to their respective blue regions, but cannot reach it, and the stable strategy is found at the maximum of FI, i.e, at the critical point. This is confirmed by our simulations, as illustrated by the top panel of Fig. \ref{fig:potential}. As a consequence, competition pushes agents to the critical region, with highest variability of internal patterns.

Some questions arise from a game-theory perspective: what is the best strategy for individuals: to cooperate, to compete, or just to be neutral? What occurs if we take the optimal strategy for any specific case an introduce a small amount of individuals with different strategies? Is the previous optimal strategy still stable? These questions require of further investigation, that we leave for a future work.

\section{\label{sec:conclusions}Discussion}
In this paper we have studied the dynamics of a community of individuals evolving to mimic other fellow agents in its community, i.e. to enhance the information each one captures from the others.  The framework employed here is similar to that in our precedent work \cite{our-PNAS}: each individual is constituted by an internal boolean network obeying an internal dynamics, which can be modulated by changing its internal parameters in the direction which maximizes its similitude to others. The model introduced here is adaptive as opposed to the one in \cite{our-PNAS} which was evolutionary; its main advantage of the model presented here is that allows for a mathematical treatment, which was mostly missing in \cite{our-PNAS}.

 As a first step, using a well established method for the study of interacting stochastic processes\cite{Dean} and through several approximations we are able to compute analytically the attractors of the dynamics, confirming our previous findings: the community self-tunes to the neighborhood of the critical point, separating ordered from disordered internal states. It is important to stress this result: the critical point can be identified with the region of parameter space with maximum variability in terms of information theory; the ``imitation game'' studied here thus leads to, surprisingly, maximum complexity. This is similar to previous results by Kaneko and Suzuki \cite{Kaneko1994}, who studied two abstract ``birds'' that play to imitate each other's ``songs''. The attractor of such dynamics is located at the ``edge of chaos'', exhibiting maximum variability \cite{Kaneko1994}.

 In the simplest version of the model introduced here, individuals adapt with the only aim of improving their individual knowledge.  As a second step, we have studied a more general scenario introducing a few basic mechanisms usually studied in game theory: cooperation (adaptation to improve the amount of information collected by each individual agent as well as by the others in the community), competition (adaptation to improve each agent's information and reduce that of the others). In these cases, we observe that, while cooperation leads the community to regions with scarce variability (i.e. to the ordered or disordered regions), criticality emerges for dominantly-competing communities.

 We hypothesize that the mechanisms reported here and their relationship with social interactions may be related with the emergence of
 criticality at a community level, as for instance with flocks of birds \cite{Bialek2014}. 
 More ambitiously, the interplay between cooperation and competition and the emergence of complex patterns in communities might be related with the large phenotypic variability observed in some bacterial communities \cite{Kussel}, a survival strategy by which individuals diversify their behavior to minimize long-term extinction risks, usually referred to as ``bet-hedging'' \cite{Ripa2009}. However, clarifying this point would require further investigation.

 Summing up, the question about whether biological organisms are poised to criticality is being harshly debated, but setting the problem from an information-theory point of view opens new and stimulating questions that seem worth exploring.

\section{\label{sec:acknowledgments}Acknowledgments}
We are grateful to J. R. Banavar for helpful discussions and valuable suggestions.  J.H. thanks R. Cerezo for constant support. M.A.M. and J.H. acknowledge support from the Spanish MINECO Excellence project FIS2013-43201-P.  S.S. thanks the support of the University of Padova, Physics and Astronomy Department (Senior Grant 129/2013 Prot. 1634).

\vspace{2cm} 

\appendix

\section{Concavity of V(D)}
\label{sec:concavity}
In this appendix we illustrate the role of the concavity/convexity of the information-deficit (pseudo)potential, $V(D(\xx,\yy))$.
For the sake of simplicity let us consider the case of a community consisting of just two individuals with parameters $\xx$ and $\yy$, respectively. Introducing the parametrization of Eq. (\ref{eq:parametrization}) in the definition of the KL divergence (Eq. (\ref{eq:KL})), it can be easily checked that:
\begin{equation}
\nabla_\xx D(\yy,\xx) = -\nabla_\yy D(\xx,\yy) = \langle \pphi \rangle_\xx - \langle \pphi \rangle_\yy.
\end{equation}
With this, we can rewrite the force $\mathbf{F}$ (Eq. (\ref{eq:drift})) as:
\begin{equation}
\mathbf{F}(\yy,\xx)=V'(D(\yy,\xx)) \left(\langle \pphi \rangle_\xx - \langle \pphi \rangle_\yy\right),
\label{eq:drift2}
\end{equation}
and consequently, the drifts exerted on the two agents have opposite directions --i.e. individuals tend to approach each other to exhibit similar patterns-- and their relative strength, $\frac{|\mathbf{F}(\yy,\xx)|}{|\mathbf{F}(\xx,\yy)|}$, is given by $\frac{V'(D(\yy,\xx))}{V'(D(\xx,\yy))}$. This means that the individual with large value of $V'(D)$ experiences a larger drift.

\begin{figure}
  \centering\includegraphics[width=0.7\columnwidth]{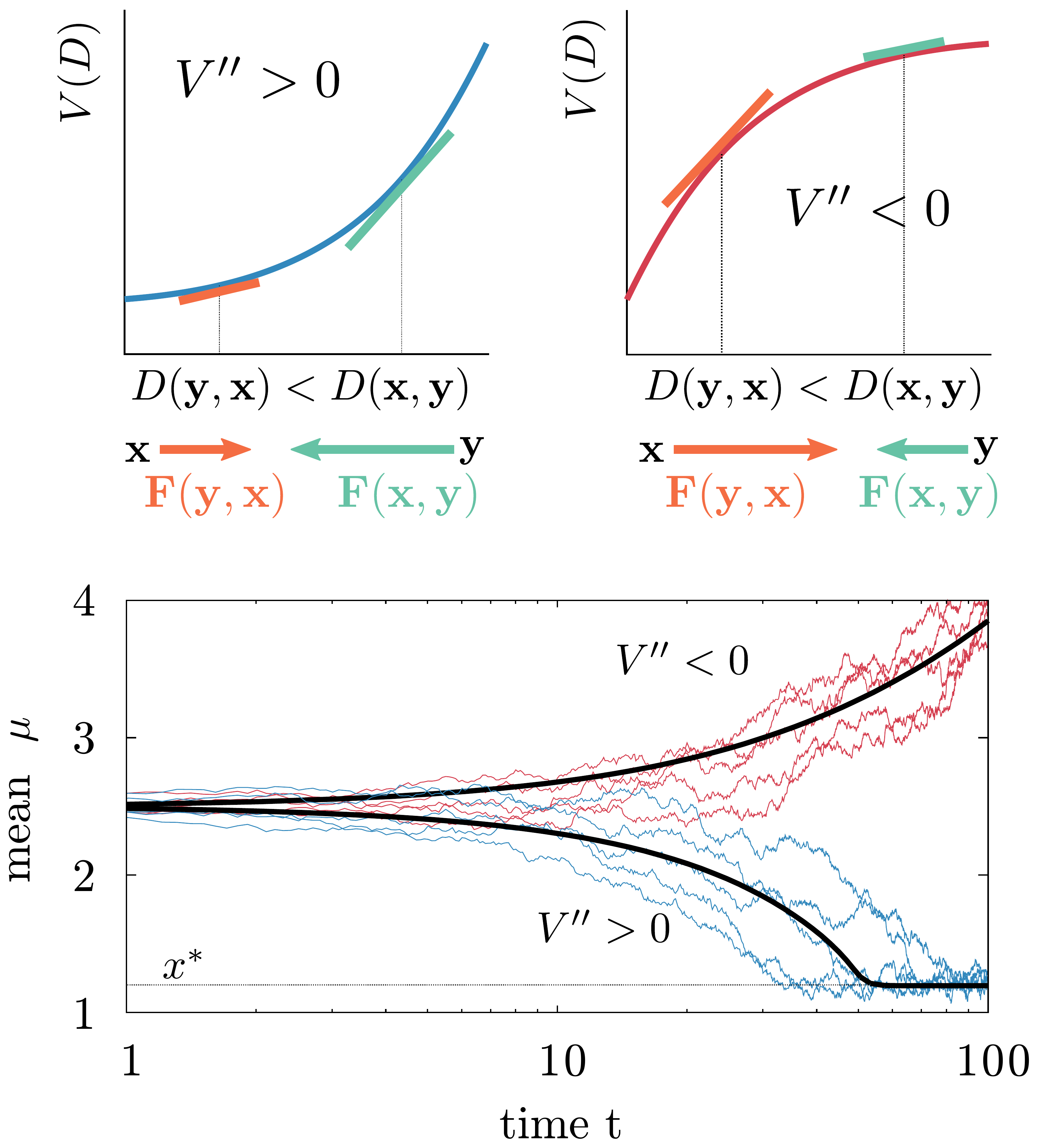} \caption{(Top panels) The relative force that two individuals with parameters $\xx$ and $\yy$, respectively, exert onto each other is given by $|\mathbf{F}(\yy,\xx)|/|\mathbf{\mathbf{F}(\xx,\yy)}|=V'(D(\yy,\xx))/V'(D(\xx,\yy))$, which means that the one with higher $V'(D)$ experiences a higher drift (derivatives have been represented with tangent lines). Consequently, if $V''>0$ (left top panel), individual with less $D$ also moves less, while the opposite occurs for $V''<0$ (right top panel), where the individual with less $D$ experiences a higher drift and is more likely to scape. (Bottom panel) The concavity/convexity of $V$ switches the dynamics. We represent the mean parameter of the community in time for choices of $V(D)$ with opposite concavity: we plot 10 independent realizations of the dynamics, Eq. (\ref{eq:langevin}), for the case (blue curves) $V(D)=\log(D+1)$ ($V''(D)=-(D+1)^2<0$) and (red curves) $V(D)=\exp(x)-1$ ($V''(D)=\exp(D)>0$), and the Ising mean-field-like parametrization, Eq. (\ref{eq:ising_1d}). While individuals approach to the critical point, $x^*$, for $V''>0$, they scape from it for $V''<0$, confirming that the sign of $V''$ inverts the dynamics. Solid black lines represent the theoretical prediction, given by the numerical integration of eqs. \ref{eq:firstmom1-1dim} and \ref{eq:secondmom1-1dim}.  The initial condition is $P(x,t=0)=\delta(x-2.5)$, and parameter values are $N=M=100$ and $T=0.1$.}
  \label{fig:concavity} \end{figure}

The forthcoming argument is illustrated in Fig. \ref{fig:concavity}. Without loss of generality, suppose that $D(\yy,\xx)<D(\xx,\yy)$. Then, in our framework, the individual with parameter $\xx$ has more information than the individual with $\yy$, and it is more fit. If $V(D)$ is a convex function ($V''>0$), as $V(D)$ is monotonically increasing, it follows that $V'(D(\yy,\xx))<V'(D(\xx,\yy))$ (see left top panel in Fig. \ref{fig:concavity}). Consequently, by Eq. (\ref{eq:drift2}), individual $\yy$ experiences a higher drift than $\xx$, and their centroid moves towards the individual with lower KL divergence, in this case $\xx$. The opposite occurs for a concave $V(D)$ (corresponding to right top panel in Fig. \ref{fig:concavity}). If $V''<0$, we can see that $V'(D(\yy,\xx))>V'(D(\xx,\yy))$, and the centroid approaches to the individual with higher KL divergence, here $\yy$.

This is confirmed by simulations of the dynamics (i.e., by numerical integration of Eq. (\ref{eq:langevin})): in bottom panel of Fig. \ref{fig:concavity} we represent 10 independent realizations for two choices of $V(D)$ (Eq. (\ref{eq:langevin})) with opposite concavity, in a community of $N=100$ individuals (blue curves correspond to $V(D)=V''>0$, and red curves to $V(D)=V''<0$). In the case of $V''>0$, the community moves toward the critical point, while it escapes from it for $V''<0$. This result is understood analytically by means of Eq. (\ref{eq:firstmom1-1dim}).

The linear case, $V(D)=D$, can be solved exactly:
\begin{equation}
\dot{\mmu} = \displaystyle \int d\xx \rho(\xx,t) \displaystyle \int d\yy \rho(\yy,t) \left(\langle \pphi\rangle_\yy -\langle \pphi \rangle_\xx \right) = 0,
\end{equation}
and, as both individuals perform equal jumps in opposite directions, the centroid does not change at all.

Summarizing, the dynamics is not completely determined with the statement that ``individuals change their parameters in the direction which maximizes their information''; \emph{one needs also to specify how much individuals move}. Under an adaptive/evolutionary framework, the more plausible scenario corresponds to the case in which fitter individuals (i.e. with more information) experience lower drifts than poorly fitted individuals (i.e. with low information), which corresponds to the choice of $V''>0$.  The opposite choice, $V''<0$, becomes senseless in our model, as it represents the case in which fittest individuals are more likely to scape to other regions in the space of parameters.  Therefore, we restrict our analysis to the choice with $V''>0$.

\section{Derivation of the equation for the evolution of the density using Dean's method} \label{sec:Dean} In this appendix we briefly review the calculation done by Dean to obtain a dynamical equation for the density of a system of interacting ``particles'' each one obeying a Langevin dynamics \cite{Dean}.  For this, we define the contribution of each individual $i$ to the global distribution of parameters as $\rho^i(\xx,t)=\delta(\xx-\xx^i(t))$, so that $\rho(\xx,t)=\sum_{i}\rho(\xx^i)/N$. Then, any function of the form $f(\xx^i)$ can be written as: \begin{equation} f(\xx^i)=\int d\xx \rho^i(\xx, t) f(\xx), \label{ref:Dean1} \end{equation} and its time derivative as: \begin{equation}
 \frac{df(\xx^i)}{dt} = \int d\xx \frac{\partial \rho^i(\xx,t)}{\partial t} f(\xx).
 \label{eq:Dean2}
\end{equation}
Alternatively, the integral can be also computed expanding the argument using the Ito's calculus; by doing so one finds:
\begin{eqnarray}
\fl \frac{d f(\xx^i)}{d t}
 &=& \int d\xx \rho^i(\xx,t) \Bigg(\nabla_\xx f(\xx) \cdot \Bigg( \sqrt{2T} \eeta^i(t) -\frac{1}{N} \sum_{j=1}^N \nabla_\xx V(D(\xx^j(t),\xx)) \Bigg)  + T\nabla_\xx^2 f (\xx) \Bigg)\nonumber\\
\fl &=& \int d\xx f(\xx) \Bigg( -\sqrt{2T} \nabla_\xx\cdot\left( \rho^i(\xx,t) \eeta^i(t) \right)+
 \left. \nabla_\xx\cdot \left( \rho^i(\xx,t) \left( \frac{1}{N} \sum_{j=1}^N \nabla_\xx V(D(\xx^j(t),\xx)) \right)\right)\right.+\nonumber\\
 \fl&&T\nabla_\xx^2 \rho^i (\xx,t) \Bigg),
\label{eq:Dean3}
 \end{eqnarray}
where we have integrated by parts at the second step. Identifying the arguments of equations \ref{eq:Dean2} and \ref{eq:Dean3}, one obtains:
 \begin{equation}
 \frac{\partial \rho^i(\xx,t)}{\partial t} = -\sqrt{2T} \nabla_\xx\cdot\left( \rho^i(\xx,t) \eeta^i(t) \right) + \nabla_\xx\cdot \left( \rho^i(\xx,t) \left( \frac{1}{N} \sum_{j=1}^N \nabla_\xx V(D(\xx^j(t),\xx)) \right)\right) + T\nabla_\xx^2 \rho^i (\xx,t),
 \end{equation}
that, averaging over individuals, transforms to:
 \begin{equation}
 \frac{\partial \rho(\xx,t)}{\partial t}= -\frac{\sqrt{2T}}{N}\sum_{i=1}^N \nabla_\xx\cdot\left( \rho^i(\xx,t) \eeta^i(t)\right) + \nabla_\xx\cdot \left( \rho(\xx,t) \left( \frac{1}{N} \sum_{j=1}^N  \nabla_\xx V(D(\xx^j(t),\xx)) \right)\right) + T\nabla_\xx^2 \rho (\xx,t).
 \end{equation}
 However, this is not a closed equation for $\rho(\xx,t)$, as it still depends on $-\frac{1}{N}\sum_{i=1}^N\nabla_\xx(\rho^i(\xx,t)\eeta^i(t))$. As Dean shows, one can simply substitute it with $N^{-1/2}\nabla_\xx(\rho^{1/2}(\xx,t)\xxi(\xx,t))$, in which $\xxi(\xx,t)$ is a new delta-correlated Gaussian noise, as it has exactly the same correlation function \cite{Dean}. After doing that, we finally recover Eq. (\ref{eq:dean}). Note that, compared to Dean's result, we have got an extra $N^{-1/2}$ factor because we have normalized the global density function to $1$ rather than $N$.

\section{$V'(0)=0$, case}
\label{sec:v0}
For the case in which $V(D)$ has a local minimum at $D=0$ (i.e., $V'(0)=0$, $V''(0)>0$), Eq. (\ref{eq:secondmom1}) vanishes, and we have to keep further terms in the expansion, obtaining a slightly more complex expression:
\begin{eqnarray} \dot K_{\alpha\beta}(t) &=& 2T\delta_{\alpha\beta}-\frac{1}{2}{V''(0)}\sum_{\gamma,\epsilon,\theta,\sigma,\tau=1}^d(\delta_{\alpha\sigma}\delta_{\beta\tau}+\delta_{\alpha\tau}\delta_{\beta\sigma}) \times \nonumber \\
  && \left(\chi_{\gamma\epsilon}(\mmu)\chi_{\theta\sigma}(\mmu)+\chi_{\gamma\theta}(\mmu)\chi_{\epsilon\sigma}(\mmu)+\chi_{\gamma\sigma}(\mmu)\chi_{\epsilon\theta}(\mmu)\right) \left(K_{\gamma\epsilon}K_{\theta\tau} + \frac{K_{\gamma\epsilon\theta\tau}}{3}\right)\nonumber\\ \label{eq:secondmom2} \end{eqnarray}

In particular, for the case of just one-parameter, $d=1$, we obtain:
\begin{equation}
 \dot K = 2T-6V''(0) \chi(\mu)^2 K^2. \label{eq:secondmom2-1dim}
\end{equation}

The stationary variance in this case is $K^*=\frac{\sqrt{T}}{\sqrt{3V''(0)}\chi(x^*)}$.

\section*{References}
\bibliographystyle{iopart-num}

\providecommand{\newblock}{}

\newpage


\setcounter{equation}{0} 
\setcounter{section}{0} 
\setcounter{figure}{0}
\setcounter{table}{0}
\setcounter{page}{1}

\begin{center}
\begin{Large}
\textbf{
Supplementary Material:\\
Cooperation, competition and the emergence of criticality in communities of adaptive systems}
\end{Large}

\hspace*{1cm}

J. Hidalgo, J. Grilli, S. Suweis, A. Maritan and  M. \'A. Mu\~noz
\end{center}

\vspace{0.5cm}

\parskip=2mm
\parindent=0mm

\renewcommand{\thesection}{S\arabic{section}}
\renewcommand{\figurename}{Supplementary Figure}
\renewcommand{\tablename}{Supplementary Table}
\renewcommand{\thefigure}{S\arabic{figure}}
\renewcommand{\thetable}{S\arabic{table}}
\renewcommand{\theequation}{S\arabic{equation}}
\renewcommand{\thepage}{S\arabic{page}}

\section{Explicit expansion}
\label{sec:appendix1}
We present the explicit calculations developed in Section IV (simple model) and Section V (cooperation/competition model) in the main text. 
As the former can be understood as a particular case of the latter (taking the symmetry coefficient $\nu=1/2$), here we only refer to the more the general case.

We start from the dynamics, given by the set of equations:
\begin{equation}
   \dot{\xx}^i = -\nabla_{\xx^i} \left( \frac{1}{N} \sum_{j\neq i}^N V^\nu (\xx^j,\xx^i)\right) +\sqrt{2T} \eeta^i(t), \quad i=1,...,N
   \label{eq:langevin-nu}
\end{equation}
where $V^\nu(\xx^j,\xx^i)=V(D(\xx^j,\xx^i))+(2\nu-1)V(D(\xx^i,\xx^j))$, $\sqrt{2T}$ modulates the noise amplitude, and $\eeta^i$ is a white noise with $\langle \eta^i_\alpha(t) \rangle=0$ and $\langle \eta^i_\alpha(t) \eta^j_\beta (t') \rangle = \delta^{ij} \delta_{\alpha\beta} \delta(t-t')$, with $i,j=1,...,N$, and $\alpha,\beta=1,...,d$.

Following Dean's method as explained in Section IVA and in Appendix B, for an infinite community, $N\rightarrow\infty$, the evolution of the probability distribution of parameters is given by: 
\begin{eqnarray}
\partial_t \rho(\xx,t) = -\nabla_\xx \cdot \left( \rho(\xx,t) \displaystyle \int d\yy \rho(\yy,t) \mathbf{F}^\nu(\yy,\xx) \right) + T\nabla^2_{\xx}\rho(\xx,t)
\label{eq:dean-nu}
\end{eqnarray}
where $\mathbf{F}^{\nu}(\yy,\xx)$ is the generalized attraction force that an individual with parameter $\yy$ produces on an individual with parameter $\xx$, defined as:
\begin{equation}
 \mathbf{F}^\nu(\yy,\xx) = -\nabla_\xx  V^\nu(\yy,\xx).
 \label{eq:drift-nu}
\end{equation}
From Eq. (\ref{eq:dean-nu}), it is easy to obtain the evolution of the mean and covariance matrix in the community multiplying by $x_\epsilon$ and $(x_\epsilon-\mu_\epsilon(t))(x_\tau-\mu_\tau(t))$, respectively, and integrating over $\xx$. After integrating by parts, one gets:
\begin{eqnarray}
\fl \dot \mu_{\epsilon} &=& \displaystyle \int d\xx \rho(\xx,t) \displaystyle \int d\yy \rho(\yy,t) F_\epsilon^\nu (\yy,\xx),
 \label{eq:mu1-nu}\\
\fl \dot K_{\epsilon\tau} &=& \sum_{\theta,\sigma=1}^d (\delta_{\theta\epsilon}\delta_{\sigma\tau}+\delta_{\theta\tau}\delta_{\sigma\epsilon})\displaystyle \int d\xx \rho(\xx,t) (x_\theta-\mu_\theta) \displaystyle \int d\yy \rho(\yy,t) F_\sigma^\nu (\yy,\xx) + 2T\delta_{\epsilon\tau}.
  \label{eq:K1-nu}
\end{eqnarray}
Let start by expanding $\mathbf{F}^\nu(\yy,\xx)$ around $(\xx,\yy) = (\mmu(t),\mmu(t))$: 
\begin{equation}
\mathbf{F}^\nu(\yy,\xx) = \sum_{n_x,n_y=0}^\infty \,\sum_{\{\aalpha\}_{n_x},\{\bbeta\}_{n_y}}\frac{\mathbf{F}^{\nu\{\aalpha\}_{n_x}\{\bbeta\}_{n_y}}}{n_x!n_y!} (\xx-\mmu)^{\{\aalpha\}_{n_x}}
(\yy-\mmu)^{\{\bbeta\}_{n_y}},
\end{equation}
where we have defined the product $(\xx-\mmu)^{\{\aalpha\}_{n_x}} :=  \prod_{i=1}^{n_x} (x_{\alpha_i}-\mu_{\alpha_i})$ and the coefficients $\mathbf{F}^{\nu\{\aalpha\}_{n_x}\{\bbeta\}_{n_y}}$ as:
\begin{equation}
\label{eq:Fdefinition}
\mathbf{F}^{\nu\{\aalpha\}_{n_x}\{\bbeta\}_{n_y}} = 
\mathbf{F}^{\nu\{\alpha_1,...,\alpha_{n_x}\}_{n_x}\{\beta_1,...,\beta_{n_y}\}_{n_y}} :=
\left.
\prod_{i=1}^{n_x} \frac{\partial}{\partial x_{\alpha_i}} \prod_{j=1}^{n_y} \frac{\partial}{\partial y_{\beta_j}} \, \mathbf{F}^\nu(\yy,\xx)
\right|_{(\xx,\yy)=(\mmu,\mmu)}.
\end{equation}
Introducing this expression in Eq. (\ref{eq:mu1-nu}), we obtain
\begin{eqnarray}
\fl\label{eq:taylor}
\dot\mu_\epsilon = &  F_\epsilon^{\nu\{\}_0\{\}_0} + \frac{1}{2!}\sum_{\alpha,\beta=1}^d\left( F_\epsilon^{\nu\{\alpha\beta\}_2\{\}_0} + F^{\nu\{\}_0\{\alpha\beta\}_2} \right) K_{\alpha\beta} +
\frac{1}{3!}\sum_{\alpha,\beta,\gamma=1}^d\left( F_\epsilon^{\nu\{\alpha\beta\gamma\}_3\{\}_0} + F_\epsilon^{\nu\{\}_0\{\alpha\beta\gamma\}_3} \right) K_{\alpha\beta\gamma}+\nonumber\\
\fl& \frac{1}{2!2!}\sum_{\alpha,\beta,\gamma,\delta=1}^d F_\epsilon^{\nu\{\alpha\beta\}_2\{\gamma\delta\}_2} K_{\alpha\beta}K_{\gamma\delta} +
\frac{1}{4!}\sum_{\alpha,\beta,\gamma,\delta=1}^d\left( F_\epsilon^{\nu\{\alpha\beta\gamma\delta\}_4\{\}_0} + F_\epsilon^{\nu\{\}_0\{\alpha\beta\gamma\delta\}_4} \right) K_{\alpha\beta\gamma\delta}+...
\end{eqnarray}
Let us note that we do not include the elements of the form $F_\epsilon^{\nu\{\alpha\}_1\{\beta_1,...,\beta_{n_y}\}_{n_y}}$ and
$F_\epsilon^{\nu\{\alpha_1,...,\alpha_{n_x}\}_{n_x}\{\beta\}_1}$ as they are multiplied by $K_{\alpha}$ and $K_{\beta}$, respectively, which are identically 0 by definition.
The first contributing terms are (see Section \ref{sec:appendix2}):
\begin{eqnarray}
\frac{1}{2}\left( F_\epsilon^{\nu\{\alpha\beta\}_2\{\}_0} + F_\epsilon^{\nu\{\}_0\{\alpha\beta\}_2} \right) &=& (1-2\nu) V'(0) \chi_{\alpha\beta\epsilon},
\end{eqnarray}
and, introducing in Eq. (\ref{eq:taylor}), we finally obtain Eq. (17) in the main text.
In the case in which $\nu=1/2$ or $V'(0)=0$, the first non-vanishing terms are (see section \ref{sec:appendix2}):
\begin{eqnarray}
\fl\frac{1}{2}\left(F_\epsilon^{\nu\{\alpha\beta\}_2\{\gamma\delta\}_2} + F_\epsilon^{\nu\{\gamma\delta\}_2\{\alpha\beta\}_2} \right)
&=&(1-\nu) V''(0) \chi_{\alpha\epsilon}(\mmu)\chi_{\beta\gamma\delta}(\mmu) +\nonumber\\
\fl&& (1-2\nu) V''(0) \chi_{\alpha\beta}(\mmu)\chi_{\gamma\delta\epsilon}(\mmu)+  \perm{\alpha,\beta,\gamma,\delta}\\
\fl F_\epsilon^{\nu\{\alpha\beta\gamma\delta\}_4\{\}_0} + F_\epsilon^{\nu\{\}_0\{\alpha\beta\gamma\delta\}_4}
&=&  4(1-2\nu) V'(0)\chi_{\alpha\beta\gamma\delta\epsilon}(\mmu) + 2(1-\nu) V''(0) \chi_{\alpha\epsilon}(\mmu)\chi_{\beta\gamma\delta}(\mmu) + \nonumber\\
&&   2(1-2\nu) V''(0)\chi_{\alpha\beta}(\mmu)\chi_{\gamma\delta\epsilon}(\mmu)+\perm{\alpha,\beta,\gamma,\delta}
\label{eq:F40_F04}
\end{eqnarray}
where $\perm{\alpha_1,...,\alpha_p}$ represents the minimal set of terms, obtained via index permutations of its precedent elements, that we have to include to symmetrize the expression under the exchange of indexes $\{\alpha_1,...,\alpha_p\}$, and where $\chi_{\alpha\beta\gamma\delta\epsilon}(\mmu)=\left.\partial_{x_\alpha}\partial_{x_\beta}\partial_{x_\gamma}\chi_{\delta\epsilon}(\xx)\right|_{\xx=\mmu}$  (see Appendix \ref{sec:appendix2}). Note that, as the ``$\{\}_2\{\}_2$'' term in Eq. (\ref{eq:taylor}) is multiplied by $K_{\alpha\beta}K_{\gamma\delta}$ and summed over
$\alpha,\beta,\gamma,\delta$, we have explicitly symmetrized it. 
For the cases of interest, i.e. $\nu=1/2$ and/or $V'(0)=0$, the term proportional to $\chi_{\alpha\beta\gamma\delta\epsilon}(\mmu)$ in Eq. (\ref{eq:F40_F04}) vanishes, and we obtain the following relation:
\begin{eqnarray}
F_\epsilon^{\nu\{\alpha\beta\}_2\{\gamma\delta\}_2} + F_\epsilon^{\nu\{\gamma\delta\}_2\{\alpha\beta\}_2} = F_\epsilon^{\nu\{\alpha\beta\gamma\delta\}_4\{\}_0} + F_\epsilon^{\nu\{\}_0\{\alpha\beta\gamma\delta\}_4}.
\end{eqnarray}
Introducing this expression in Eq. (\ref{eq:taylor}), we finally obtain
\begin{eqnarray}
\fl \dot \mu_\epsilon(t) & = & (1-\nu)\frac{V''(0)}{4} \sum_{\alpha,\beta,\gamma,\delta=1}^d \left( \chi_{\alpha\epsilon}(\mmu)\chi_{\beta\gamma\delta}(\mmu) 
+ \perm{\alpha,\beta,\gamma,\delta} \right) \left(K_{\alpha\beta}K_{\gamma\delta} + \frac{K_{\alpha\beta\gamma\delta}}{3} \right) + \nonumber \\
\fl && (1-2\nu)\frac{V''(0)}{4}\sum_{\alpha,\beta,\gamma,\delta=1}^d \left(  \chi_{\alpha\beta}(\mmu) \chi_{\gamma\delta\epsilon}(\mmu)
+ \perm{\alpha,\beta,\gamma,\delta} \right) \left(K_{\alpha\beta}K_{\gamma\delta} + \frac{K_{\alpha\beta\gamma\delta}}{3} \right)\nonumber\\
\fl&&\label{eq:firstmom2-nu}
\end{eqnarray}
that, for $\nu=1/2$, it reduces to Eq. (12) in the main text.

Now we proceed analogously for the expansion of Eq. (\ref{eq:K1-nu}):
\begin{eqnarray}
\label{eq:taylor2}
\fl \dot K_{\epsilon\tau} &=&  2T\delta_{\epsilon\tau} +
\sum_{\theta,\sigma=1}^d (\delta_{\theta\epsilon}\delta_{\sigma\tau}+\delta_{\theta\tau}\delta_{\sigma\epsilon}) \left( \sum_{\alpha=1}^d  F_\sigma^{\nu\{\alpha\}_1\{\}_0} K_{\theta\alpha}   
 +\frac{1}{2!}\sum_{\alpha,\beta=1}^d F_\sigma^{\nu\{\alpha\beta\}_2\{\}_0} K_{\theta\alpha\beta} + \right.\nonumber\\
\fl &&\left. \frac{1}{3!} \sum_{\alpha,\beta,\gamma=1}^d  F_\sigma^{\nu\{\alpha\beta\gamma\}_3\{\}_0} K_{\theta\alpha\beta\gamma}  +  \frac{1}{1!2!} \sum_{\alpha,\beta,\gamma=1}^d F_\sigma^{\nu\{\alpha\}_1\{\beta\gamma\}_2}  K_{\theta\alpha}  K_{\beta\gamma} \right)+...
\end{eqnarray}
where we have omitted the terms multiplied by $K_{\theta}$, which is identically zero by definition.

If $V'(0)\neq 0$, we can just take the first term in the expansion of Eq. (\ref{eq:taylor2}), and then we arrive to Eq. (18) in the main text. 
In the case $V'(0)=0$ or $\nu=0$, the whole first row of Eq. (\ref{eq:taylor2}) vanishes and we have next-order contributions. For $V'(0)=0$ and $\nu\neq0$ we obtain:
\begin{eqnarray}
\fl \dot K_{\epsilon\tau}(t) & = & 2T\delta_{\epsilon\tau} -\nu{V''(0)}\sum_{\alpha,\beta,\gamma=1}^d\left(\chi_{\alpha\beta}(\mmu)\chi_{\gamma\epsilon}(\mmu)+\perm{\alpha,\beta,\gamma,\epsilon}\right) \left(K_{\alpha\beta}K_{\gamma\tau} + \frac{K_{\alpha\beta\gamma\tau}}{3}\right)+\nonumber\\
\fl &&-\nu{V''(0)}\sum_{\alpha,\beta,\gamma=1}^d\left(\chi_{\alpha\beta}(\mmu)\chi_{\gamma\tau}(\mmu)+\perm{\alpha,\beta,\gamma,\tau}\right) 
\left(K_{\alpha\beta}K_{\gamma\epsilon} + \frac{K_{\alpha\beta\gamma\epsilon}}{3}\right)
\label{eq:secondmom2-nu}
\end{eqnarray}
while, for the case $V'(0)\neq0$ and $\nu=0$, we have:
\begin{eqnarray}
 \dot K_{\epsilon\tau}(t) & = & 2T\delta_{\epsilon\tau} + \frac{V'(0)}{3} \sum_{\alpha,\beta,\gamma=1}^d\left(\chi_{\alpha\beta\gamma\epsilon}(\mmu) K_{\alpha\beta\gamma\tau} + \chi_{\alpha\beta\gamma\tau}(\mmu) K_{\alpha\beta\gamma\epsilon}\right),
\end{eqnarray}
where $\chi_{\alpha\beta\gamma\epsilon}(\mmu)=\left.\partial_{x_\alpha}\partial_{x_\beta}\chi_{\gamma\epsilon}(\xx)\right|_{\xx=\mmu}$ (see Section \ref{sec:appendix2}).

\section{Derivatives of the force $\mathbf{F}^\nu$}
\label{sec:appendix2}
Here we give the expressions for the derivatives of the generalized potential used in the previous section.
To simplify the nomenclature, we introduce, for any function $f=f(\yy,\xx)$ (as for instance $V^\nu(\yy,\xx)=V(D(\yy,\xx))+(2\nu-1)V(D(\xx,\yy))$), the notation:
\begin{equation}
 \crossd{f}{\alpha_1...\alpha_{n_x}}{n_x}{\beta_1...\beta_{n_y}}{n_y} := \left.\prod_{i=1}^{n_x}\frac{\partial}{\partial x_{\alpha_i}} \prod_{j=1}^{n_y}\frac{\partial}{\partial y_{\beta_j}}
f(\yy,\xx)\right|_{(\xx,\yy)=(\mmu,\mmu)}.
\end{equation}
From this definition, it follows that
\begin{equation}
 \crossd{F^\nu_\epsilon}{\alpha_1...\alpha_{n_x}}{n_x}{\beta_1...\beta_{n_y}}{n_y} = -\left(\crossd{V}{\alpha_1...\alpha_{n_x}\epsilon}{n_x+1}{\beta_1...\beta_{n_y}}{n_y}+(2\nu-1)\crossd{V}{\beta_1...\beta_{n_y}}{n_y}{\alpha_1...\alpha_{n_x}\epsilon}{n_x+1	}\right).
\end{equation}

Let start by computing the first derivatives of the FI in terms of the observables functions, $\pphi$. Using the parametrization of Eq. (3) in the main text, 
we have:
\begin{eqnarray}
\fl \chi_{\alpha\beta}(\xx) & = -\partial_{x_\alpha}\langle\phi_\beta\rangle_\xx & = \langle \left[\langle \phi_\alpha \rangle_\xx - \phi_\alpha\right]\left[\langle \phi_\beta \rangle_\xx -
\phi_\beta\right] \rangle_\xx\\
\fl \chi_{\alpha\beta\gamma}(\xx) & = \partial_{x_\alpha} \chi_{\beta\gamma}(\xx) & = \langle \left[\langle \phi_\alpha \rangle_\xx - \phi_\alpha\right]\left[\langle \phi_\beta \rangle_\xx -
\phi_\beta\right]\left[\langle \phi_\gamma \rangle_\xx - \phi_\gamma\right] \rangle_\xx\\
\fl \chi_{\alpha\beta\gamma\delta}(\xx) & = \partial_{x_\alpha} \chi_{\beta\gamma\delta}(\xx) & = \langle \left[\langle \phi_\alpha \rangle_\xx - \phi_\alpha\right]\left[\langle
\phi_\beta \rangle_\xx - \phi_\beta\right] \left[\langle \phi_\gamma \rangle_\xx - \phi_\gamma\right] \left[\langle \phi_\delta \rangle_\xx - \phi_\delta\right] \rangle_\xx \nonumber\\
\fl & & \qquad - \chi_{\alpha\beta}(\xx)\chi_{\gamma\delta}(\xx) + \perm{\alpha,\beta,\gamma,\delta}\\
\fl \chi_{\alpha\beta\gamma\delta\epsilon}(\xx) & = \partial_{x_\alpha} \chi_{\beta\gamma\delta\epsilon}(\xx) & = \langle \left[\langle \phi_\alpha \rangle_\xx - \phi_\alpha\right]\left[\langle
\phi_\beta \rangle_\xx - \phi_\beta\right] \left[\langle \phi_\gamma \rangle_\xx - \phi_\gamma\right] \left[\langle \phi_\delta \rangle_\xx - \phi_\delta\right] \left[\langle \phi_\epsilon \rangle_\xx
- \phi_\epsilon\right] \rangle_\xx \nonumber\\
\fl & & \qquad - \chi_{\alpha\beta}(\xx)\chi_{\gamma\delta\epsilon}(\xx) + \perm{\alpha,\beta,\gamma,\delta,\epsilon}
\end{eqnarray}
where $\perm{\alpha_1,...,\alpha_p}$ represents the minimal set of terms, obtained via index permutations of its precedent elements, that we have to include to symmetrize the expression under the exchange of indexes $\{\alpha_1,...,\alpha_p\}$. Let us note that each of the definitions above is symmetric under the exchange of any two indexes. Now, we compute the cross partial derivatives of the KL divergence, $D(\yy,\xx)$, that are used in the next step: 

\begin{eqnarray}
\crossd{D}{\alpha}{1}{}{0} & = & 0\\
\crossd{D}{\alpha\beta}{2}{}{0} & = & \chi_{\alpha\beta}(\mmu)\\
\crossd{D}{\alpha\beta\gamma}{3}{}{0} & = & \chi_{\alpha\beta\gamma}(\mmu)\\
\crossd{D}{\alpha\beta\gamma\delta}{4}{}{0} & = & \chi_{\alpha\beta\gamma\delta}(\mmu)\\
\crossd{D}{\alpha\beta\gamma\delta\epsilon}{5}{}{0} & = & \chi_{\alpha\beta\gamma\delta\epsilon}(\mmu)
\end{eqnarray}\begin{eqnarray}
\crossd{D}{}{0}{\alpha}{1} & = & 0\\
\crossd{D}{}{0}{\alpha\beta}{2} & = & \chi_{\alpha\beta}(\mmu)\\
\crossd{D}{}{0}{\alpha\beta\gamma}{3} & = & 2\chi_{\alpha\beta\gamma}(\mmu)\\
\crossd{D}{}{0}{\alpha\beta\gamma\delta}{4} & = & 3\chi_{\alpha\beta\gamma\delta}(\mmu)\\
\crossd{D}{}{0}{\alpha\beta\gamma\delta\epsilon}{5} & = & 4\chi_{\alpha\beta\gamma\delta\epsilon}(\mmu)
\end{eqnarray}\begin{eqnarray}
\crossd{D}{\alpha}{1}{\beta}{1} & = & -\chi_{\alpha\beta}(\mmu)\\
\crossd{D}{\alpha}{1}{\beta\gamma}{2} & = & -\chi_{\alpha\beta\gamma}(\mmu)\\
\crossd{D}{\alpha}{1}{\beta\gamma\delta}{3} & = & -\chi_{\alpha\beta\gamma\delta}(\mmu)\\
\crossd{D}{\alpha}{1}{\beta\gamma\delta\epsilon}{4} & = & -\chi_{\alpha\beta\gamma\delta\epsilon}(\mmu)
\end{eqnarray}\begin{eqnarray}
\crossd{D}{\alpha\beta}{2}{\gamma}{1} & = &  0\\
\crossd{D}{\alpha\beta\gamma}{3}{\delta}{1} & = &  0\\
\crossd{D}{\alpha\beta\gamma\delta}{4}{\epsilon}{1} & = & 0\\
\crossd{D}{\alpha\beta}{2}{\gamma\delta}{2} & = & 0\\
\crossd{D}{\alpha\beta\gamma}{3}{\delta\epsilon}{2} & = & 0\\
\crossd{D}{\alpha\beta}{2}{\gamma\delta\epsilon}{3} & = & 0
\end{eqnarray}

Finally, expression of the terms contributing in the expansion of Eqs. (\ref{eq:taylor}) and (\ref{eq:taylor2}) are listed below (non-vanishing cross derivatives of $D$ are marked in \enh{blue}):
\begin{itemize}

\small
\item $\crossd{F_\epsilon^\nu}{}{0}{}{0}$:
\begin{eqnarray}
\fl \crossd{V}{\epsilon}{1}{}{0} & = & V'(0) \crossd{D}{\epsilon}{1}{}{0} = 0\\
\fl \crossd{V}{}{0}{\epsilon}{1} & = & V'(0) \crossd{D}{}{0}{\epsilon}{1} = 0\\
\fl \nonumber\\
\fl \crossd{F_\epsilon^\nu}{}{0}{}{0} & = & -\left(\crossd{V}{\epsilon}{1}{}{0} + (2\nu-1)\crossd{V}{}{0}{\epsilon}{1}\right) = 0
\end{eqnarray}

\item $\crossd{F_\epsilon^\nu}{\alpha}{1}{}{0}$:
\begin{eqnarray}
\fl \crossd{V}{\alpha\epsilon}{2}{}{0} &=& V''(0) \crossd{D}{\alpha}{1}{}{0} \crossd{D}{\epsilon}{1}{}{0} + V'(0) \enh{ \crossd{D}{\alpha\epsilon}{2}{}{0} } = V'(0) \chi_{\alpha\epsilon}(\mmu)\\
\fl \crossd{V}{}{0}{\alpha\epsilon}{2} &=& V''(0) \crossd{D}{}{0}{\alpha}{1} \crossd{D}{}{0}{\epsilon}{1} + V'(0) \enh{ \crossd{D}{}{0}{\alpha\epsilon}{2} } = V'(0) \chi_{\alpha\epsilon}(\mmu)\\
\fl \nonumber\\
\fl \crossd{F_\epsilon^\nu}{\alpha}{1}{}{0} &=& -\left(\crossd{V}{\alpha\epsilon}{2}{}{0}+(2\nu-1)\crossd{V}{}{0}{\alpha\epsilon}{2} \right) = -2 \nu V'(0) \chi_{\alpha\epsilon}(\mmu)
\end{eqnarray}

\item $\crossd{F_\epsilon^\nu}{\alpha\beta}{2}{}{0}$:
\begin{eqnarray}
\fl \crossd{V}{\alpha\beta\epsilon}{3}{}{0} & = & V^{(3)}(0) \crossd{D}{\alpha}{1}{}{0} \crossd{D}{\beta}{1}{}{0} \crossd{D}{\epsilon}{1}{}{0} + V''(0) \crossd{D}{\alpha\beta}{2}{}{0} \crossd{D}{\epsilon}{1}{}{0} + \nonumber\\
\fl && V'(0) \enh{ \crossd{D}{\alpha\beta\epsilon}{3}{}{0} } + \perm{\alpha,\beta,\epsilon} = V'(0)\chi_{\alpha\beta\epsilon}(\mmu)\\
\fl \crossd{V}{}{0}{\alpha\beta\epsilon}{3} & = & V^{(3)}(0) \crossd{D}{}{0}{\alpha}{1} \crossd{D}{}{0}{\beta}{1} \crossd{D}{}{0}{\epsilon}{1} + V''(0) \crossd{D}{}{0}{\alpha\beta}{2} \crossd{D}{}{0}{\epsilon}{1} + \nonumber\\
\fl && V'(0) \enh{ \crossd{D}{}{0}{\alpha\beta\epsilon}{3} } + \perm{\alpha,\beta,\epsilon} = 2V'(0)\chi_{\alpha\beta\epsilon}(\mmu)\\
\fl \nonumber\\
\fl \crossd{F_\epsilon^\nu}{\alpha\beta}{2}{}{0} &=& -\left( \crossd{V}{\alpha\beta\epsilon}{3}{}{0} +(2\nu-1)\crossd{V}{}{0}{\alpha\beta\epsilon}{3} \right) = (1-4\nu)V'(0) \chi_{\alpha\beta\epsilon}(\mmu)
\end{eqnarray}

\item $\crossd{F_\epsilon^\nu}{}{0}{\alpha\beta}{2}$:
\begin{eqnarray}
\fl \crossd{V}{\epsilon}{1}{\alpha\beta}{2} &=& V^{(3)}(0) \crossd{D}{}{0}{\alpha}{1} \crossd{D}{}{0}{\beta}{1} \crossd{D}{\epsilon}{1}{}{0} + V''(0) \left( \crossd{D}{\epsilon}{1}{\alpha}{1} \crossd{D}{}{0}{\beta}{1} + \crossd{D}{}{0}{\alpha\beta}{2} \crossd{D}{\epsilon}{1}{}{0} \right) + \nonumber\\
\fl & & V'(0) \enh{ \crossd{D}{\epsilon}{1}{\alpha\beta}{2} } + \perm{\alpha,\beta} = -V'(0)\chi_{\alpha\beta\epsilon}(\mmu)\\
\fl \crossd{V}{\alpha\beta}{2}{\epsilon}{1} &=& V^{(3)}(0) \crossd{D}{\alpha}{1}{}{0} \crossd{D}{\beta}{1}{}{0} \crossd{D}{}{0}{\epsilon}{1} + V''(0) \left( \crossd{D}{\alpha}{1}{\epsilon}{1} \crossd{D}{\beta}{1}{}{0} + \crossd{D}{\alpha\beta}{2}{}{0} \crossd{D}{}{0}{\epsilon}{1} \right) + \nonumber\\
\fl & & V'(0) \crossd{D}{\alpha\beta}{2}{\epsilon}{1} + \perm{\alpha,\beta} = 0\\
\fl \nonumber\\
\fl \crossd{F_\epsilon^\nu}{}{0}{\alpha\beta}{2} &=& -\left( \crossd{V}{\epsilon}{1}{\alpha\beta}{2} + (2\nu-1)\crossd{V}{\alpha\beta}{2}{\epsilon}{1} \right) = V'(0) \chi_{\alpha\beta\epsilon}(\mmu)
\end{eqnarray}

\item $\crossd{F_\epsilon^\nu}{\alpha\beta\gamma}{3}{}{0}$:
\begin{eqnarray}
\fl \crossd{V}{\alpha\beta\gamma\epsilon}{4}{}{0} & = & V^{(4)}(0) \crossd{D}{\alpha}{1}{}{0} \crossd{D}{\beta}{1}{}{0} \crossd{D}{\gamma}{1}{}{0} \crossd{D}{\epsilon}{1}{}{0} + V^{(3)}(0) \crossd{D}{\alpha\beta}{2}{}{0} \crossd{D}{\gamma}{1}{}{0} \crossd{D}{\epsilon}{1}{}{0} + \nonumber\\
\fl & & V''(0) \left( \crossd{D}{\alpha\beta\gamma}{3}{}{0} \crossd{D}{\epsilon}{1}{}{0} + \enh{ \crossd{D}{\alpha\beta}{2}{}{0} \crossd{D}{\gamma\epsilon}{2}{}{0} } \right) + V'(0) \enh{ \crossd{D}{\alpha\beta\gamma\epsilon}{4}{}{0} } + \perm{\alpha,\beta,\gamma,\epsilon} + \nonumber\\
\fl &=& V'(0)\chi_{\alpha\beta\gamma\epsilon}(\mmu) + V''(0)\chi_{\alpha\beta}(\mmu)\chi_{\gamma\epsilon}(\mmu) + \perm{\alpha,\beta,\gamma,\epsilon}\\
\fl \crossdinv{V}{\alpha\beta\gamma\epsilon}{4}{}{0} & = & V^{(4)}(0) \crossdinv{D}{\alpha}{1}{}{0} \crossdinv{D}{\beta}{1}{}{0} \crossdinv{D}{\gamma}{1}{}{0} \crossdinv{D}{\epsilon}{1}{}{0} + V^{(3)}(0) \crossdinv{D}{\alpha\beta}{2}{}{0} \crossdinv{D}{\gamma}{1}{}{0} \crossdinv{D}{\epsilon}{1}{}{0} + \nonumber\\
\fl & & V''(0) \left( \crossdinv{D}{\alpha\beta\gamma}{3}{}{0} \crossdinv{D}{\epsilon}{1}{}{0} + \enh{ \crossdinv{D}{\alpha\beta}{2}{}{0} \crossdinv{D}{\gamma\epsilon}{2}{}{0} } \right) + V'(0) \enh{ \crossdinv{D}{\alpha\beta\gamma\epsilon}{4}{}{0} } + \perm{\alpha,\beta,\gamma,\epsilon} + \nonumber\\
\fl &=& 3V'(0)\chi_{\alpha\beta\gamma\epsilon}(\mmu) + V''(0)\chi_{\alpha\beta}(\mmu)\chi_{\gamma\epsilon}(\mmu) + \perm{\alpha,\beta,\gamma,\epsilon}\\
\fl \nonumber\\
\fl \crossd{F_\epsilon^\nu}{\alpha\beta\gamma}{3}{}{0} & = & -\left( \crossd{V}{\alpha\beta\gamma\epsilon}{4}{}{0} +(2\nu-1)\crossdinv{V}{\alpha\beta\gamma\epsilon}{4}{}{0} \right) =
 2(1-3\nu)V'(0)\chi_{\alpha\beta\gamma\epsilon}(\mmu) \nonumber\\
\fl &&-2\nu V''(0)\chi_{\alpha\beta}(\mmu)\chi_{\gamma\epsilon}(\mmu) + \perm{\alpha,\beta,\gamma,\epsilon}
\end{eqnarray}

\item $\crossd{F_\epsilon^\nu}{}{0}{\alpha\beta\gamma}{3}$:
\begin{eqnarray}
\fl \crossd{V}{\epsilon}{1}{\alpha\beta\gamma}{3} & = & V^{(4)}(0) \crossd{D}{}{0}{\alpha}{1} \crossd{D}{}{0}{\beta}{1} \crossd{D}{}{0}{\gamma}{1}
\crossd{D}{\epsilon}{1}{}{0} + \nonumber\\
\fl & & V^{(3)}(0) \left(\crossd{D}{}{0}{\alpha\beta}{2} \crossd{D}{}{0}{\gamma}{1} \crossd{D}{\epsilon}{1}{}{0} + \crossd{D}{\alpha}{1}{\epsilon}{1}
\crossd{D}{}{0}{\beta}{1} \crossd{D}{}{0}{\gamma}{1} \right ) + \nonumber\\
\fl & & V''(0) \left(\crossd{D}{}{0}{\alpha\beta\gamma}{3} \crossd{D}{\epsilon}{1}{}{0} + \crossd{D}{\epsilon}{1}{\alpha\beta}{2} \crossd{D}{}{0}{\gamma}{1} +\enh{  \crossd{D}{}{0}{\alpha\beta}{2} \crossd{D}{\epsilon}{1}{\gamma}{1} } \right) + \nonumber\\
\fl & & V'(0) \enh{ \crossd{D}{\epsilon}{1}{\alpha\beta\gamma}{3} } + \perm{\alpha,\beta,\gamma} \nonumber\\
\fl & = & -V'(0)\chi_{\alpha\beta\gamma\epsilon}(\mmu) - V''(0)\chi_{\alpha\beta}(\mmu)\chi_{\gamma\epsilon}(\mmu) +\perm{\alpha,\beta,\gamma,\epsilon}\\
\fl \crossdinv{V}{\epsilon}{1}{\alpha\beta\gamma}{3} & = & V^{(4)}(0) \crossdinv{D}{}{0}{\alpha}{1} \crossdinv{D}{}{0}{\beta}{1} \crossdinv{D}{}{0}{\gamma}{1}
\crossdinv{D}{\epsilon}{1}{}{0} + \nonumber\\
\fl & & V^{(3)}(0) \left(\crossdinv{D}{}{0}{\alpha\beta}{2} \crossdinv{D}{}{0}{\gamma}{1} \crossdinv{D}{\epsilon}{1}{}{0} + \crossdinv{D}{\alpha}{1}{\epsilon}{1}
\crossdinv{D}{}{0}{\beta}{1} \crossdinv{D}{}{0}{\gamma}{1} \right ) + \nonumber\\
\fl & & V''(0) \left(\crossdinv{D}{}{0}{\alpha\beta\gamma}{3} \crossdinv{D}{\epsilon}{1}{}{0} + \crossdinv{D}{\epsilon}{1}{\alpha\beta}{2} \crossdinv{D}{}{0}{\gamma}{1} +\enh{  \crossdinv{D}{}{0}{\alpha\beta}{2} \crossdinv{D}{\epsilon}{1}{\gamma}{1} } \right) + \nonumber\\
\fl & & V'(0) \crossdinv{D}{\epsilon}{1}{\alpha\beta\gamma}{3} + \perm{\alpha,\beta,\gamma} = - V''(0)\chi_{\alpha\beta}(\mmu)\chi_{\gamma\epsilon}(\mmu) +\perm{\alpha,\beta,\gamma,\epsilon}\nonumber\\
\\
\fl \crossd{F_\epsilon^\nu}{}{0}{\alpha\beta\gamma}{3} &=&-\left(\crossd{V}{\epsilon}{1}{\alpha\beta\gamma}{3}+(2\nu-1)\crossdinv{V}{\epsilon}{1}{\alpha\beta\gamma}{3}\right)= V'(0)\chi_{\alpha\beta\gamma\epsilon}(\mmu)\nonumber\\
\fl &&+2\nu V''(0)\chi_{\alpha\beta}(\mmu)\chi_{\gamma\epsilon}(\mmu) +\perm{\alpha,\beta,\gamma,\epsilon}
\end{eqnarray}

\item $\crossd{F_\epsilon^\nu}{\alpha\beta}{2}{\gamma\delta}{2}$:

\begin{eqnarray}
\fl \crossd{V}{\alpha\beta\epsilon}{2}{\gamma\delta}{2}&=&
V^{(5)}(0) \crossd{D}{\alpha}{1}{}{0} \crossd{D}{\beta}{1}{}{0} \crossd{D}{}{0}{\gamma}{1} \crossd{D}{}{0}{\delta}{1} \crossd{D}{\epsilon}{1}{}{0} + \nonumber\\
\fl & & V^{(4)}(0)\left(
\crossd{D}{\alpha\beta}{2}{}{0} \crossd{D}{}{0}{\gamma}{1} \crossd{D}{}{0}{\delta}{1} \crossd{D}{\epsilon}{1}{}{0}
+ \crossd{D}{\alpha}{1}{}{0}\crossd{D}{\beta}{1}{}{0}\crossd{D}{}{0}{\gamma\delta}{2}\crossd{D}{\epsilon}{1}{}{0} \right. + \nonumber\\
\fl & & \left. \qquad\qquad \crossd{D}{\alpha}{1}{\gamma}{1} \crossd{D}{\beta}{1}{}{0}\crossd{D}{}{0}{\delta}{1}\crossd{D}{\epsilon}{1}{}{0} \right) + \nonumber\\
\fl & & V^{(3)}(0)\left( \crossd{D}{\alpha\beta\epsilon}{3}{}{0}\crossd{D}{}{0}{\gamma}{1}\crossd{D}{}{0}{\delta}{1}
+ \crossd{D}{\alpha\beta}{2}{\gamma}{1}\crossd{D}{}{0}{\delta}{1}\crossd{D}{\epsilon}{1}{}{0} + \right.\nonumber\\
\fl & & \left. \qquad\qquad\crossd{D}{\alpha}{1}{\gamma\delta}{2}\crossd{D}{}{0}{\beta}{1}\crossd{D}{\epsilon}{1}{}{0} + \crossd{D}{\alpha\beta}{2}{}{0} \crossd{D}{}{0}{\gamma\delta}{2}
\crossd{D}{\epsilon}{1}{}{0} + \right. \nonumber\\
\fl & & \left. \qquad\qquad
\crossd{D}{\alpha}{1}{\gamma}{1}\crossd{D}{\beta}{1}{\delta}{1}\crossd{D}{\epsilon}{1}{}{0} + 
\crossd{D}{\alpha\beta}{2}{}{0}\crossd{D}{\epsilon}{1}{\gamma}{1}\crossd{D}{}{0}{\delta}{1} \right) +
\nonumber\\
\fl & & V''(0) \left( \crossd{D}{\alpha\beta}{2}{\gamma\delta}{2} \crossd{D}{\epsilon}{1}{}{0} + \crossd{D}{\alpha\beta\epsilon}{3}{\gamma}{1} \crossd{D}{}{0}{\delta}{1}
 + \enh{ \crossd{D}{\alpha\beta\epsilon}{3}{}{0} \crossd{D}{}{0}{\gamma\delta}{2} } \right. + \nonumber\\
\fl & & \left.\qquad\qquad  \crossd{D}{\alpha\beta}{2}{\gamma}{1} \crossd{D}{\epsilon}{1}{\delta}{1} + \enh{ \crossd{D}{\alpha}{1}{\gamma\delta}{2} \crossd{D}{\beta\epsilon}{2}{}{0} }\right) + \nonumber\\
\fl & & V'(0) \crossd{D}{\alpha\beta\epsilon}{3}{\gamma\delta}{2} + \perm{\alpha,\beta,\epsilon][\gamma,\delta} \nonumber \\
\fl & = & V''(0)\left( \chi_{\alpha\beta\epsilon}(\mmu)\chi_{\gamma\delta}(\mmu) - \chi_{\alpha\gamma\delta}(\mmu)\chi_{\beta\epsilon}(\mmu) \right) +
\perm{\alpha,\beta,\epsilon][\gamma\delta}
\end{eqnarray}\begin{eqnarray}
\fl \crossdinv{V}{\alpha\beta\epsilon}{3}{\gamma\delta}{2}&=&
V^{(5)}(0) \crossdinv{D}{\alpha}{1}{}{0} \crossdinv{D}{\beta}{1}{}{0} \crossdinv{D}{}{0}{\gamma}{1} \crossdinv{D}{}{0}{\delta}{1} \crossdinv{D}{\epsilon}{1}{}{0} + \nonumber\\
\fl & & V^{(4)}(0)\left(
\crossdinv{D}{\alpha\beta}{2}{}{0} \crossdinv{D}{}{0}{\gamma}{1} \crossdinv{D}{}{0}{\delta}{1} \crossdinv{D}{\epsilon}{1}{}{0}
+ \crossdinv{D}{\alpha}{1}{}{0}\crossdinv{D}{\beta}{1}{}{0}\crossdinv{D}{}{0}{\gamma\delta}{2}\crossdinv{D}{\epsilon}{1}{}{0} \right. + \nonumber\\
\fl & & \left. \qquad\qquad \crossdinv{D}{\alpha}{1}{\gamma}{1} \crossdinv{D}{\beta}{1}{}{0}\crossdinv{D}{}{0}{\delta}{1}\crossdinv{D}{\epsilon}{1}{}{0} \right) + \nonumber\\
\fl & & V^{(3)}(0)\left( \crossdinv{D}{\alpha\beta\epsilon}{3}{}{0}\crossdinv{D}{}{0}{\gamma}{1}\crossdinv{D}{}{0}{\delta}{1}
+ \crossdinv{D}{\alpha\beta}{2}{\gamma}{1}\crossdinv{D}{}{0}{\delta}{1}\crossdinv{D}{\epsilon}{1}{}{0} + \right.\nonumber\\
\fl & & \left. \qquad\qquad\crossdinv{D}{\alpha}{1}{\gamma\delta}{2}\crossdinv{D}{}{0}{\beta}{1}\crossdinv{D}{\epsilon}{1}{}{0} + \crossdinv{D}{\alpha\beta}{2}{}{0} \crossdinv{D}{}{0}{\gamma\delta}{2}
\crossdinv{D}{\epsilon}{1}{}{0} + \right. \nonumber\\
\fl & & \left. \qquad\qquad
\crossdinv{D}{\alpha}{1}{\gamma}{1}\crossdinv{D}{\beta}{1}{\delta}{1}\crossdinv{D}{\epsilon}{1}{}{0} + 
\crossdinv{D}{\alpha\beta}{2}{}{0}\crossdinv{D}{\epsilon}{1}{\gamma}{1}\crossdinv{D}{}{0}{\delta}{1} \right) +
\nonumber\\
\fl & & V''(0) \left( \crossdinv{D}{\alpha\beta}{2}{\gamma\delta}{2} \crossdinv{D}{\epsilon}{1}{}{0} + \crossdinv{D}{\alpha\beta\epsilon}{3}{\gamma}{1} \crossdinv{D}{}{0}{\delta}{1}
 + \enh{ \crossdinv{D}{\alpha\beta\epsilon}{3}{}{0} \crossdinv{D}{}{0}{\gamma\delta}{2} } \right. + \nonumber\\
\fl & & \left.\qquad\qquad  \enh{\crossdinv{D}{\alpha\beta}{2}{\gamma}{1} \crossdinv{D}{\epsilon}{1}{\delta}{1}} + \crossdinv{D}{\alpha}{1}{\gamma\delta}{2} \crossdinv{D}{\beta\epsilon}{2}{}{0} \right) + \nonumber\\
\fl & & V'(0) \crossdinv{D}{\alpha\beta\epsilon}{3}{\gamma\delta}{2} + \perm{\alpha,\beta,\epsilon][\gamma,\delta} \nonumber \\
\fl &=&V''(0)\left(2\chi_{\alpha\beta\epsilon}(\mmu)\chi_{\gamma\delta}(\mmu)+\chi_{\alpha\beta\gamma}(\mmu)\chi_{\delta\epsilon}(\mmu)\right)+\perm{\alpha,\beta,\epsilon][\gamma,\delta}\\
\fl \nonumber\\
\fl \crossd{F_\epsilon^\nu}{\alpha\beta}{2}{\gamma\delta}{2} &=&-\left(\crossd{V}{\alpha\beta\epsilon}{3}{\gamma\delta}{2}+(2\nu-1)\crossdinv{V}{\alpha\beta\epsilon}{3}{\gamma\delta}{2}\right)=V''(0) \chi_{\alpha\beta\gamma}(\mmu)\chi_{\delta\epsilon}(\mmu)+\perm{\alpha,\beta,\gamma,\delta,\epsilon}\nonumber\\
\fl && -2\nu V''(0)\left(2\chi_{\alpha\beta\epsilon}(\mmu)\chi_{\gamma\delta}(\mmu)+\chi_{\alpha\beta\gamma}(\mmu)\chi_{\delta\epsilon}(\mmu)\right)+\perm{\alpha,\beta,\epsilon][\gamma,\delta}
\end{eqnarray}

\item $\crossd{F_\epsilon^\nu}{\alpha\beta\gamma\delta}{4}{}{0}$:

\begin{eqnarray}
\fl \crossd{V}{\alpha\beta\gamma\delta\epsilon}{5}{}{0} & = & V^{(5)}(0) \crossd{D}{\alpha}{1}{}{0}\crossd{D}{\beta}{1}{}{0}\crossd{D}{\gamma}{1}{}{0}\crossd{D}{\delta}{1}{}{0}\crossd{D}{\epsilon}{1}{}{0}
+ \nonumber\\
\fl & & V^{(4)}(0) \crossd{D}{\alpha\beta}{2}{}{0}\crossd{D}{\gamma}{1}{}{0}\crossd{D}{\delta}{1}{}{0}\crossd{D}{\epsilon}{1}{}{0}+\nonumber\\
\fl & & V^{(3)}(0) \left( \crossd{D}{\alpha\beta\gamma}{3}{}{0}\crossd{D}{\delta}{1}{}{0}\crossd{D}{\epsilon}{1}{}{0} +
\crossd{D}{\alpha\beta}{2}{}{0}\crossd{D}{\gamma\delta}{2}{}{0}\crossd{D}{\epsilon}{1}{}{0} \right) +\nonumber\\
\fl & & V''(0) \left( \crossd{D}{\alpha\beta\gamma\delta}{4}{}{0}\crossd{D}{\epsilon}{1}{}{0} + \enh{ \crossd{D}{\alpha\beta\gamma}{3}{}{0}\crossd{D}{\delta\epsilon}{2}{}{0} } \right) + \nonumber\\
\fl & & V'(0) \enh{ \crossd{D}{\alpha\beta\gamma\delta\epsilon}{5}{}{0} } + \perm{\alpha,\beta,\gamma,\delta,\epsilon} \nonumber \\
\fl & = &
V'(0)\chi_{\alpha\beta\gamma\delta\epsilon}(\mmu) + V''(0) \chi_{\alpha\beta\gamma}(\mmu)\chi_{\delta\epsilon}(\mmu) + \perm{\alpha,\beta,\gamma,\delta,\epsilon}\\
\fl \crossdinv{V}{\alpha\beta\gamma\delta\epsilon}{5}{}{0} & = & V^{(5)}(0) \crossdinv{D}{\alpha}{1}{}{0}\crossdinv{D}{\beta}{1}{}{0}\crossdinv{D}{\gamma}{1}{}{0}\crossdinv{D}{\delta}{1}{}{0}\crossdinv{D}{\epsilon}{1}{}{0}
+ \nonumber\\
\fl & & V^{(4)}(0) \crossdinv{D}{\alpha\beta}{2}{}{0}\crossdinv{D}{\gamma}{1}{}{0}\crossdinv{D}{\delta}{1}{}{0}\crossdinv{D}{\epsilon}{1}{}{0}+\nonumber\\
\fl & & V^{(3)}(0) \left( \crossdinv{D}{\alpha\beta\gamma}{3}{}{0}\crossdinv{D}{\delta}{1}{}{0}\crossdinv{D}{\epsilon}{1}{}{0} +
\crossdinv{D}{\alpha\beta}{2}{}{0}\crossdinv{D}{\gamma\delta}{2}{}{0}\crossdinv{D}{\epsilon}{1}{}{0} \right) +\nonumber\\
\fl & & V''(0) \left( \crossdinv{D}{\alpha\beta\gamma\delta}{4}{}{0}\crossdinv{D}{\epsilon}{1}{}{0} + \enh{ \crossdinv{D}{\alpha\beta\gamma}{3}{}{0}\crossdinv{D}{\delta\epsilon}{2}{}{0} } \right) + \nonumber\\
\fl & & V'(0) \enh{ \crossdinv{D}{\alpha\beta\gamma\delta\epsilon}{5}{}{0} } + \perm{\alpha,\beta,\gamma,\delta,\epsilon} \nonumber \\
\fl & = & 4V'(0)\chi_{\alpha\beta\gamma\delta\epsilon}(\mmu)+2V''(0)\chi_{\alpha\beta\gamma}(\mmu)\chi_{\delta\epsilon}(\mmu) + \perm{\alpha,\beta,\gamma,\delta,\epsilon}\\
\fl \nonumber\\
\fl \crossd{F_\epsilon^\nu}{\alpha\beta\gamma\delta}{4}{}{0} &=&-\left(\crossd{V}{\alpha\beta\gamma\epsilon}{5}{}{0}+(2\nu-1)\crossdinv{V}{}{0}{\alpha\beta\gamma\delta\epsilon}{5}\right)=\nonumber\\
\fl &=&(3-8\nu)V'(0)\chi_{\alpha\beta\gamma\delta\epsilon}(\mmu)+(1-4\nu)V''(0)\chi_{\alpha\beta\gamma}(\mmu)\chi_{\delta\epsilon}(\mmu) + \perm{\alpha,\beta,\gamma,\delta,\epsilon}\nonumber\\
\end{eqnarray}

\item $\crossd{F_\epsilon^\nu}{}{0}{\alpha\beta\gamma\delta}{4}$:
\begin{eqnarray}
\fl \crossd{V}{\epsilon}{1}{\alpha\beta\gamma\delta}{4} & = & V^{(5)}(0) \crossd{D}{}{0}{\alpha}{1}\crossd{D}{}{0}{\beta}{1}\crossd{D}{}{0}{\gamma}{1}\crossd{D}{}{0}{\delta}{1}\crossd{D}{\epsilon}{0}{}{0}
+ \nonumber \\
\fl & & V^{(4)}(0) \left(
\crossd{D}{}{0}{\alpha\beta}{2}\crossd{D}{}{0}{\gamma}{1}\crossd{D}{}{0}{\delta}{1}\crossd{D}{\epsilon}{0}{}{0} +
\crossd{D}{\epsilon}{1}{\alpha}{1}\crossd{D}{}{0}{\beta}{1}\crossd{D}{}{0}{\gamma}{1}\crossd{D}{}{0}{\delta}{1}
 \right) + \nonumber\\
\fl & & V^{(3)}(0) \left( 
\crossd{D}{}{0}{\alpha\beta\gamma}{3}\crossd{D}{}{0}{\delta}{1}\crossd{D}{\epsilon}{1}{}{0} + 
\crossd{D}{\epsilon}{1}{\alpha\beta}{2}\crossd{D}{\gamma}{0}{\delta}{1}\crossd{D}{}{0}{\delta}{1}
 \right. + \nonumber\\
\fl & & \qquad\qquad \left. \crossd{D}{}{0}{\alpha\beta}{2}\crossd{D}{}{0}{\gamma\delta}{2}\crossd{D}{\epsilon}{1}{}{0} +
\crossd{D}{\epsilon}{1}{\alpha}{1}\crossd{D}{}{0}{\beta\gamma}{2}\crossd{D}{}{0}{\delta}{1} \right) + \nonumber\\
\fl & & V''(0) \left( \crossd{D}{}{0}{\alpha\beta\gamma\delta}{4}\crossd{D}{\epsilon}{1}{}{0} + \crossd{D}{\epsilon}{1}{\alpha\beta\gamma}{3}\crossd{D}{}{0}{\delta}{1}\right. + \nonumber \\
\fl & & \qquad\qquad\left. \enh{ \crossd{D}{\epsilon}{1}{\alpha\beta}{2}\crossd{D}{}{0}{\gamma\delta}{2} } + \enh{ \crossd{D}{}{0}{\alpha\beta\gamma}{3}\crossd{D}{\epsilon}{1}{\delta}{1} } \right) +
\nonumber\\
\fl & & V'(0) \enh{ \crossd{D}{\epsilon}{1}{\alpha\beta\gamma\delta}{4} } + \perm{\alpha,\beta,\gamma,\delta} \nonumber \\
\fl & = &
-V'(0)\chi_{\alpha\beta\gamma\delta\epsilon}(\mmu) -V''(0)\left( \chi_{\alpha\beta\epsilon}(\mmu)\chi_{\gamma\delta}(\mmu) + 2\chi_{\alpha\beta\gamma}(\mmu)\chi_{\delta\epsilon}(\mmu)
\right) + \perm{\alpha,\beta,\gamma,\delta} \nonumber\\
\fl & = & -V'(0)\chi_{\alpha\beta\gamma\delta\epsilon}(\mmu) -V''(0) \chi_{\alpha\beta\gamma}(\mmu)\chi_{\delta\epsilon}(\mmu) + \perm{\alpha,\beta,\gamma,\delta,\epsilon} \nonumber\\
\fl & &  - V''(0) \chi_{\alpha\beta\gamma}(\mmu) \chi_{\delta\epsilon} + \perm{\alpha,\beta,\gamma,\delta}\\
\fl \crossdinv{V}{\epsilon}{1}{\alpha\beta\gamma\delta}{4} & = & V^{(5)}(0) \crossdinv{D}{}{0}{\alpha}{1}\crossdinv{D}{}{0}{\beta}{1}\crossdinv{D}{}{0}{\gamma}{1}\crossdinv{D}{}{0}{\delta}{1}\crossdinv{D}{\epsilon}{0}{}{0}
+ \nonumber \\
\fl & & V^{(4)}(0) \left(
\crossdinv{D}{}{0}{\alpha\beta}{2}\crossdinv{D}{}{0}{\gamma}{1}\crossdinv{D}{}{0}{\delta}{1}\crossdinv{D}{\epsilon}{0}{}{0} +
\crossdinv{D}{\epsilon}{1}{\alpha}{1}\crossdinv{D}{}{0}{\beta}{1}\crossdinv{D}{}{0}{\gamma}{1}\crossdinv{D}{}{0}{\delta}{1}
 \right) + \nonumber\\
\fl & & V^{(3)}(0) \left( 
\crossdinv{D}{}{0}{\alpha\beta\gamma}{3}\crossdinv{D}{}{0}{\delta}{1}\crossdinv{D}{\epsilon}{1}{}{0} + 
\crossdinv{D}{\epsilon}{1}{\alpha\beta}{2}\crossdinv{D}{\gamma}{0}{\delta}{1}\crossdinv{D}{}{0}{\delta}{1}
 \right. + \nonumber\\
\fl & & \qquad\qquad \left. \crossdinv{D}{}{0}{\alpha\beta}{2}\crossdinv{D}{}{0}{\gamma\delta}{2}\crossdinv{D}{\epsilon}{1}{}{0} +
\crossdinv{D}{\epsilon}{1}{\alpha}{1}\crossdinv{D}{}{0}{\beta\gamma}{2}\crossdinv{D}{}{0}{\delta}{1} \right) + \nonumber\\
\fl & & V''(0) \left( \crossdinv{D}{}{0}{\alpha\beta\gamma\delta}{4}\crossdinv{D}{\epsilon}{1}{}{0} + \crossdinv{D}{\epsilon}{1}{\alpha\beta\gamma}{3}\crossdinv{D}{}{0}{\delta}{1}\right. + \nonumber \\
\fl & & \qquad\qquad\left. \crossdinv{D}{\epsilon}{1}{\alpha\beta}{2}\crossdinv{D}{}{0}{\gamma\delta}{2}  + \enh{ \crossdinv{D}{}{0}{\alpha\beta\gamma}{3}\crossdinv{D}{\epsilon}{1}{\delta}{1} } \right) +
\nonumber\\
\fl & & V'(0) \crossdinv{D}{\epsilon}{1}{\alpha\beta\gamma\delta}{4} + \perm{\alpha,\beta,\gamma,\delta} \nonumber \\
\fl & = & -V''(0) \chi_{\alpha\beta\gamma}(\mmu)\chi_{\delta\epsilon}(\mmu)+\perm{\alpha,\beta,\gamma,\delta} \\
\fl \nonumber\\
\fl \crossd{F_\epsilon^\nu}{}{0}{\alpha\beta\gamma\delta}{4} &=&-\left(\crossd{V}{\epsilon}{1}{\alpha\beta\gamma\delta}{4}+(2\nu-1)\crossdinv{V}{\epsilon}{1}{\alpha\beta\gamma\delta}{4}\right)=V'(0) \chi_{\alpha\beta\gamma\delta\epsilon}(\mmu) + \nonumber\\
\fl & & V''(0) \chi_{\alpha\beta\gamma}(\mmu)\chi_{\delta\epsilon}(\mmu) + \perm{\alpha,\beta,\gamma,\delta,\epsilon} + 2\nu V''(0) \chi_{\alpha\beta\gamma}(\mmu)\chi_{\delta\epsilon}(\mmu) + \perm{\alpha,\beta,\gamma,\delta}\nonumber\\
\end{eqnarray}

\item $\crossd{F_\epsilon^\nu}{\alpha}{1}{\beta\gamma}{2}$:

\begin{eqnarray}
\fl \crossd{V}{\alpha\epsilon}{2}{\beta\gamma}{2} & = & V^{(4)}(0)\crossd{D}{\alpha}{1}{}{0} \crossd{D}{}{0}{\beta}{1} \crossd{D}{}{0}{\gamma}{1} \crossd{D}{\epsilon}{1}{}{0} + \nonumber\\
\fl & & V^{(3)}(0) \left( \crossd{D}{\alpha\epsilon}{2}{}{0} \crossd{D}{}{0}{\beta}{1} \crossd{D}{}{0}{\gamma}{1} + \crossd{D}{\alpha}{1}{\beta}{1} \crossd{D}{}{0}{\gamma}{1} \crossd{D}{\epsilon}{1}{}{0} + 
\right.\nonumber\\
&& \qquad \left.
\crossd{D}{}{0}{\beta\gamma}{2} \crossd{D}{\alpha}{1}{}{0} \crossd{D}{\epsilon}{1}{}{0} \right) + \nonumber\\
\fl && V''(0) \left(\crossd{D}{\alpha\epsilon}{2}{\beta}{1} \crossd{D}{}{0}{\gamma}{1} + \crossd{D}{\alpha}{1}{\beta\gamma}{2} \crossd{D}{\epsilon}{1}{}{0} + \right. \nonumber\\
\fl && \qquad \left.
\enh{ \crossd{D}{\alpha\epsilon}{2}{}{0} \crossd{D}{}{0}{\beta\gamma}{2} } + \enh{ \crossd{D}{\alpha}{1}{\beta}{1} \crossd{D}{\epsilon}{1}{\gamma}{1} } \right)  +  V'(0) \crossd{D}{\alpha\epsilon}{2}{\beta\gamma}{2} + \perm{\alpha,\epsilon][\beta,\gamma} \nonumber \\
\fl & = & V''(0) \chi_{\alpha\beta}(\mmu)\chi_{\gamma\epsilon}(\mmu) + \perm{\alpha,\beta,\gamma,\epsilon}\\
\fl \nonumber\\
\fl \crossd{F_\epsilon^\nu}{\alpha}{1}{\beta\gamma}{2}&=&-\left(\crossd{V}{\alpha\epsilon}{2}{\beta\gamma}{2}+(2\nu-1)\crossdinv{V}{\alpha\epsilon}{2}{\beta\gamma}{2}\right) = -2\nu V''(0) \chi_{\alpha\beta}(\mmu)\chi_{\gamma\epsilon}(\mmu) +  \perm{\alpha,\beta,\gamma,\epsilon}\nonumber\\
\end{eqnarray}

\end{itemize}

\end{document}